\documentclass[12pt, aps,prd,onecolumn,showpacs,floats,floatfix,letterpaper,nofootinbib,notitlepage]{revtex4-1}
\pdfoutput=1
\usepackage{hyperref}
\usepackage{amssymb,amsmath,latexsym,mathrsfs}
\usepackage{graphicx,subfigure}
\usepackage{epsfig}
\usepackage{varioref,xr-hyper}
\usepackage{color}
\usepackage{multirow}
\usepackage{array}
\usepackage{tcolorbox}

\hypersetup{
     colorlinks   = true,
     citecolor    = violet,
     linkcolor = cyan
     }

\newcommand\Smallstrut{\rule{0pt}{2ex}}

\begin{document}

\title{Towards statistically homogeneous and isotropic perfect fluid universes with cosmic backreaction}
\author{S. M. Koksbang}
\email{sofie.koksbang@helsinki.fi}
\affiliation{Department of Physics, University of Helsinki and Helsinki Institute of Physics, P.O. Box 64, FIN-00014 University of Helsinki, Finland}

\begin{abstract}
A method for constructing statistically homogeneous and isotropic perfect fluid universe models with significant cosmic backreaction is proposed. The method is illustrated using a simplified model constructed as a Swiss-cheese model with Lemaitre-Tolman-Bondi structures. The model exhibits significant cosmic backreaction and is used to study methods proposed in the literature for relating volume averaged quantities with observations. The comparison shows a poor agreement between exact redshift-distance relations and the relations predicted by schemes based on volume averages. Most of these deviations are, however, clearly exaggerated by peculiarities of the example model, such as large local expansion rates.
\end{abstract}
\maketitle

\section{Introduction}\label{sec:intro}
Standard cosmology is based on the assumption that there exists a single Friedmann-Lemaitre-Robertson-Walker (FLRW) model that gives a good description of the spatially averaged universe both in terms of its dynamics and energy content. However, the average of a generic inhomogeneous universe has an evolution that deviates from that of an FLRW model. The effect of inhomogeneities on average evolution is known as cosmic backreaction and is most commonly described using the Buchert averaging scheme \cite{fluid1,bc_fluidII} (see e.g. \cite{fluid_generalized1,fluid_generalized2} for augmentations and \cite{bc_review1,bc_review2,also_2region} for reviews). This scheme prescribes a method for describing the large-scale/``average" dynamics of the Universe by introducing spatial averages of scalar quantities. The simplest setup for using this scheme considers spacetimes containing irrotational dust and a possible non-vanishing cosmological constant such that the spacetime can be foliated orthogonally to the fluid flow with the line element reading $ds^2 = -dt^2 + g_{ij}dx^idx^j$ ($c$ is set equal to 1, the Einstein summation convention is used and Latin letters are used as space indices while Greek letters will be used as spacetime indices). Then, the spatial average of a scalar quantity $S$ on a spatial domain $D$ is defined as $\left\langle S\right\rangle:=\frac{\int_D S\sqrt{|\det g_{ij}|}d^3x}{\int_D  \sqrt{|\det g_{ij}|}d^3x}$. Using this definition to average the Hamiltonian constraint and Raychaudhuri equation leads to evolution equations very reminiscent of the Friedmann equations (subscripted commas indicate partial derivatives and $G_N$ is Newton's constant):
\begin{equation}\label{eq:Friedmann}
\frac{1}{3}\left\langle \Theta\right\rangle^2 = 3\left( \frac{ a_{D,t}}{a_D}\right) ^2 =:3H_D^2= 8\pi G_N\left\langle \rho\right\rangle - \frac{1}{2}\left\langle ^{(3)}R\right\rangle +\Lambda - \frac{1}{2}Q
\end{equation}
\begin{equation}\label{eq:acc}
3\frac{a_{D,tt}}{a_D} = -4\pi G\left\langle \rho\right\rangle + \Lambda + Q.
\end{equation}
The (normalized) average volume scale factor $a_D$ is defined as $a_D:=\left( \frac{V_D}{V_{D_0}}\right)^{1/3} $, where $V_D$ is the volume of the domain $D$ and subscripted zeros indicate evaluation at present time. As seen, there is an extra source term, $Q$, compared to the Friedmann equations. This term, the kinematical backreaction, is defined as $Q:=\frac{2}{3}\left(\left\langle \Theta ^2\right\rangle-\left\langle \Theta\right\rangle^2 \right) -2\left\langle \sigma^2\right\rangle $, where $\Theta$ is the local expansion rate of the fluid and $\sigma^2:=\frac{1}{2}\sigma_{\mu\nu}\sigma^{\mu\nu}$ is its shear scalar. Besides the kinematical backreaction term, the averaged Hamiltonian constraint deviates from the Friedmann equation by permitting $\left\langle ^{(3)}R\right\rangle$ to evolve differently than proportional to $a_D^{-2}$. In fact, the evolution of $\left\langle ^{(3)}R\right\rangle$ is linked to the evolution of $Q$ by the integrability condition, $\frac{1}{a_D^6}\partial_t\left( a_D^6Q\right) +\frac{1}{a_D^2} \partial_t\left( a_D^2\left\langle ^{(3)}R\right\rangle\right)  =0 $, which must be fulfilled in order for equation \eqref{eq:Friedmann} to be the integral of equation \eqref{eq:acc}. This equation shows that $\left\langle ^{(3)}R\right\rangle\propto a_D^{-2}$ when $Q = 0$. In this case, the resulting average evolution is that of an FLRW model. However, it has been shown that the FLRW models are globally unstable
\footnote{The unstable nature of FLRW backgrounds is also indicated by gradient expansions as seen by the results presented in e.g. \cite{gradient}.}
\cite{unstable} and hence even a numerically small kinematical backreaction can propel the average evolution of a cosmological model away from that of the FLRW model corresponding to its ``initial" average. In such a situation, one has $Q\approx 0$ and hence $a_D^2\left\langle ^{(3)}R\right\rangle\approx$ const., and the average of the model evolves slowly between different near-FLRW states as explained in e.g. \cite{peak}. This implies that the spatial average of a model can evolve from being flat to being curved i.e. averaging can lead to the ``emergence" of curvature as recently discussed in e.g. \cite{emergence}. This effect has been shown explicitly to occur in approximate models as in e.g. \cite{virialisation,curvature_Newtonian}, in a Swiss-cheese Lemaitre-Tolman-Bondi (LTB, \cite{LTB1,LTB2,LTB3}) model \cite{Tardis}, statistical models \cite{peak,peak2} and the ``simplified" silent universes of \cite{simsilun} as well as semi-locally in Szekeres (\cite{Szekeres}) models \cite{emergence_Szekeres}.
\newline\indent
There is, of course, also the possibility that $Q$ is numerically large. In this case it might be anticipated that the resulting evolution would deviate significantly from FLRW evolutions but this is not necessarily so. Indeed, $Q$ could for instance mimic a cosmological constant with the resulting average evolution mimicking that of a $\Lambda$CDM model although this situation seems unlikely (see e.g. \cite{bc_as_lamdba_unlikely}).
\newline\newline
Cosmic backreaction is particularly interesting because it in principle has the potential to explain the apparent accelerated expansion of the Universe without introducing any exotic dark energy component as well as possibly being able to mimic dark matter \cite{BC_as_DM}. Less ambitiously, cosmic backreaction might solve the $H_0$-problem through the emergence of curvature \cite{emergence_H0}, or a small backreaction may bias the values obtained from analyses of data based on FLRW models and must therefore be identified and taken into account in an era of precision cosmology. Yet another option is that cosmic backreaction is entirely negligible in the real universe. Whichever is the case, a theoretical quantification of cosmic backreaction is necessary for getting the foundations of cosmology onto solid ground; the mathematics clearly shows that {\em in principle} backreaction terms affect the overall dynamics of the Universe. It is therefore an important goal of cosmologists to obtain a theoretical understanding of the size of cosmic backreaction in the real universe similarly to e.g. the desire to theoretically understand the value of the vacuum energy density. An important step towards reaching this goal is understanding what type of averaging scheme should actually be used: While the Buchert averaging scheme is the dominant averaging scheme used when considering backreaction, it is only relevant if the resulting volume averaged quantities can be related to observables in a meaningful way.
\newline\indent
Through theoretical considerations it was in \cite{av_obs1,av_obs2} asserted that redshift and distance measures can be described through volume averaged quantities plus statistical fluctuations in a universe that is spatially statistically homogeneous and isotropic with structures evolving slowly compared to the time it takes a light ray to traverse the homogeneity scale. However, studies based on Swiss-cheese models explicitly show that some of the assertions leading to this conclusion are not valid in general \cite{Tardis,scSZ5}. The models studied in \cite{scSZ5} had negligible backreaction and the redshift-distance relation was still well described by volume averaged quantities. In \cite{Tardis} however, neither of these were the case, but the model studied in \cite{Tardis} contained surface layers which were found to significantly affect light propagation, implying an uncertainty in the validity of the results of \cite{Tardis}. The results of \cite{Tardis} must nonetheless be carefully considered as they could be reflections of the non-negligible backreaction of the model. Specifically, it may be that the results of \cite{Tardis} indicate that another relation between volume averages and observables is more appropriate than the one suggested in \cite{av_obs1,av_obs2}. Other such relations have been suggested in the literature, with the approach of \cite{template1,template2} being particularly noteworthy as it has a thorough mathematical justification.
\newline\indent
The most trustworthy way of determining a relation between volume averaged quantities and observations is to construct exact inhomogeneous solutions to the Einstein equation that are statistically homogeneous and isotropic and which have reasonably small and slowly evolving structures but which exhibit non-negligible backreaction. The work presented here is a step towards the goal of constructing such models without introducing pathologies such as surface layers or shell crossings that can impair light propagation studies. Specifically, a scheme for constructing such models will be proposed in section \ref{sec:swisscheese} and the idea will be illustrated with a simple example model. The particular example model is merely meant as an illustration of the principles of the proposed scheme and cannot be considered realistic and can hence not be used e.g. for quantifying backreaction in a realistic setting. However, the model has no actual pathologies that can impair light propagation studies. Redshift-distance relations will therefore be studied in the model although the results should be considered with caution as peculiarities of the model may affect light propagation significantly. Theoretical aspects of exact and average light propagation are considered in sections \ref{sec:exact_light} and \ref{sec:av_light} while results from a light propagation study based on the example model are presented in section \ref{sec:results}. A summary is given in section \ref{sec:summary}.

\section{LTB Swiss-cheese models with backreaction}\label{sec:swisscheese}
Swiss-cheese models are constructed by removing spatially spherical regions of FLRW models (the ``cheese") and smoothly joining the boundary of the removed region with an inhomogeneous solution to the Einstein equation (the ``holes" in the cheese), typically containing a single inhomogeneity such as a central black hole or a mass-compensated void. The resulting Swiss-cheese model is an exact solution to the Einstein equation if the Darmois junction conditions are fulfilled \cite{Darmois}. These conditions require the metric and the extrinsic curvature to be continuous on the boundaries between the holes and the cheese. If the holes in the FLRW model are placed and oriented
\footnote{Although the removed FLRW patch is typically spherical, the structure replacing the patch does not necessarily have to be spherically symmetric. For instance, the non-spherically symmetric Szekeres models can be smoothly joined with FLRW models. See e.g. \cite{scSZ5} for examples.}
randomly, the resulting spacetime will be (spatially) statistically homogeneous and isotropic.
\newline\indent
Swiss-cheese models were first introduced in \cite{EinsteinStraus1,EinsteinStraus2} where FLRW models were joined with the Schwarzschild metric. More recently, Kottler/Schwarzschild-de Sitter metrics \cite{scKottler1,scKottler2}, Szekeres models \cite{scSZ1,scSZ2,scSZ3,scSZ4,scSZ5} and especially LTB models \cite{scLTB1,scLTB2,scLTB3,scLTB4,scLTB5,scLTB6,scLTB7,scLTB8,scLTB9,scLTB10,scLTB11,scLTB12,scLTB13,scLTB14,scLTB15} have been combined with FLRW spacetimes, usually with the purpose of understanding how inhomogeneities affect light propagation, but mostly in models with negligible backreaction or involving extremely large structures.
\newline\indent
The Swiss-cheese model constructed here will be based on an LTB model with dust and a cosmological constant.
\newline\newline
 The line element of an LTB model can be written as
\begin{equation}
\begin{split}
ds^{2} = -dt^2 +\frac{A^2_{,r}(t,r)}{1-k(r)}dr^2 +A(t,r)^2d\Omega^2.
\end{split}
\end{equation}
The evolution of $A$ is determined by the equation
\begin{equation}\label{eq:A}
A_{,t}^2 = \frac{2M}{A} - k +\frac{1}{3}\Lambda A^2,
\end{equation}
while the density is given by
\begin{equation}
\rho = \frac{2M_{,r} }{\beta A^2A_{,r} },\, \, \, \, \, \, \, \, \beta = 8\pi G_N.
\end{equation}
The equation for $A$ can be solved by simple integration, i.e.
\begin{equation}
\int_{0}^{A(t,r)}\frac{d\tilde A}{\sqrt{\frac{2M}{\Smallstrut\tilde A}-k+\frac{1}{3}\Lambda \tilde A^2}} = t-t_{bb}(r).
\end{equation}
The function $t_{bb}(r)$ represents the local time of the big bang: $A(t_{bb}(r),r) = 0$.
\newline\newline
To construct a specific LTB model two functions must be specified and in addition a rescaling of $r$ may be used to fix a third function. The following two subsections discuss how to specify an LTB model in a manner permitting significant cosmic backreaction.

\subsection{Obtaining backreaction}\label{subsec:obtainingbc}
The authors of \cite{Tardis} introduce a ``Swiss-cheese theorem" which gives conditions under which there cannot be significant backreaction in Swiss-cheese models based on dust Szekeres solutions. Since the LTB model is the spherically symmetric limit of the quasi-spherical Szekeres model, this theorem also applies to dust LTB models. The five conditions are:
\newline\newline
1) $A(t,0) = 0$
\newline\newline
2) $A_{,r}\geq 0$
\newline\newline
3) There are no singularities between $t = t_{bb}(r)$ and $t = t_0$ with $0\leq t_{bb}(r)<<t_0$.
\newline\newline
4) The spacetime matches smoothly to an FLRW dust background at a finite $r = r_b$.
\newline\newline
5) $A(t_0,r_b)$ is small compared to the background spacetime curvature radius.
\newline\newline
The proof of the theorem is based on showing that the conditions imply that $|k|<<1$ which then trivially implies that the volume of a Szekeres structure is approximately equal to the volume of the removed FLRW patch at all times (with volumes computed of spherical regions centered at $r=0$). Physically, the result is due to the fact that a local LTB/Szekeres structure with a large curvature parameter will either collapse within a short time or expand fast leading to shell crossings in the outer layers of the central structure.
\newline\indent
In \cite{Tardis}, condition 2 was broken in order to obtain significant cosmic backreaction. As mentioned in the introduction, this lead to the appearance of surface layers, violating the Darmois junction conditions and severely impacting light propagation. Therefore, another approach is proposed here based on the following consideration. From the proof of the Swiss-cheese theorem, it is clear that in order to obtain significant backreaction, the curvature parameter must be large and the problem with this is the appearance of shell crossings or very early crunch times i.e. the problem with large curvature is that dust models have no pressure to stabilize structures when densities become very large. A solution to this problem is to include an energy component which contains pressure and hence can stabilize the structure formation. The situation can be illustrated by considering FLRW models: An overdense dust model has a finite crunch time, but if a component with pressure is added, a loitering or coasting phase can appear and collapse can be avoided completely \cite{loitering}. To utilize this idea to construct a phenomenologically realistic Swiss-cheese model would require constructing a model which contains a component with an equation of state parameter and/or energy density that varies in space and/or time so that these can be chosen to locally suppress the formation of shell crossings and/or early crunch times by mimicking virialization, but at the same time vanishing outside the structure so that the cheese is effectively free of exotic components. This should be possible by using the models of \cite{mikko} (see e.g. also \cite{sussman_a_la_mp}) or variations of such models or of e.g. the Lemaitre model \cite{LTB1}. Here, in order to illustrate the idea without introducing the complexity of the models in \cite{LTB1,mikko}, only the well-known LTB model with dust and a cosmological constant will be considered. Specifically, the cosmological constant will be used to stabilize a positively curved spacetime, similar to how it can lead to a loitering phase in an overdense FLRW model. This will lead to a Swiss-cheese model that has backreaction but since the cosmological constant is spatially homogeneous, the global curvature of the entire Swiss-cheese model needs to be extremely large in order for backreaction to be significant. This implies that the modeled universe is quite small and in addition the entire Swiss-cheese model will have exponential expansion already a few Gyr after the big bang. The exponential expansion leads the inhomogeneities of the model to be spatially very large which can affect the relation between exact and average light propagation relations. However, despite these unrealistic features, the model has no actual pathologies such as surface layers, shell crossings or non-big bang singularities.

\subsection{An explicit model}\label{subsec:example}
This subsection details the construction of an LTB Swiss-cheese model with non-negligible backreaction obtained by modifying a loitering FLRW model. The model violates point 5 and the second part of point 3 of the Swiss-cheese theorem as well as by adding a cosmological constant.
\newline\newline
The LTB model considered here is based on an FLRW model specified by $\Omega_{m,1} = 16000$, $\Omega_{\Lambda,1} = 7845.91355$ and $H_1 = 70$km/s/Mpc. The subscript $``1"$ is used to indicate evaluation at $a = 1$ which here does not correspond to observation time (which is why the usual subscript $``0"$ is not used). Notice that a high level of fine-tuning is necessary when using large density parameters and an appreciable loitering phase is desired before the exponential expansion begins. Note also that the large positive curvature implies that the considered universe is quite small. This is not too important for the present work as light propagation over long distances can be facilitated simply by letting the light ray circumnavigate the modeled universe several times.
\begin{figure}
\centering
\includegraphics[scale = 0.6]{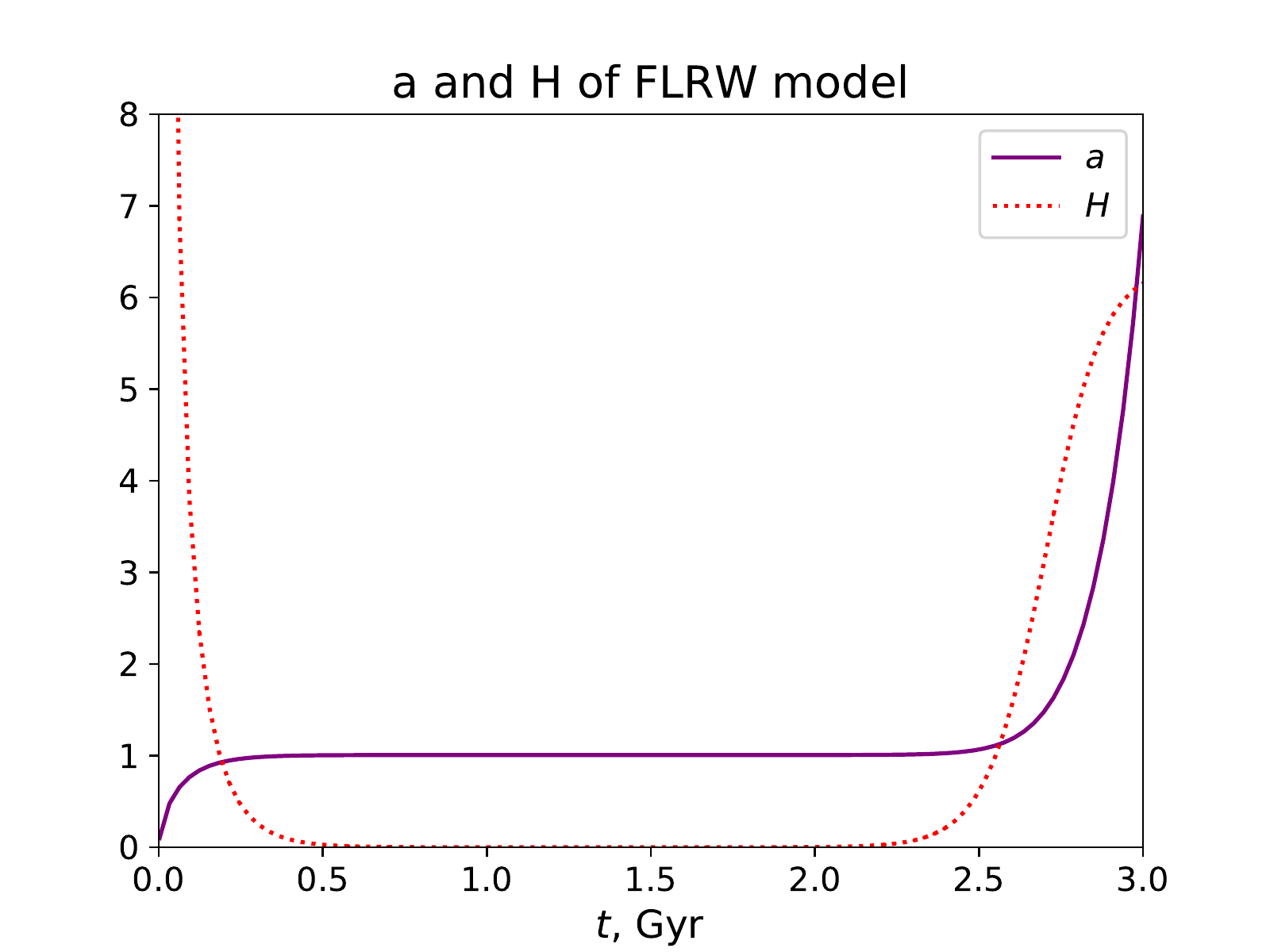}
\caption{Scale factor and Hubble rate of the considered FLRW model.}
\label{fig:a}
\end{figure}
The scale factor and Hubble rate of the model are shown in figure \ref{fig:a}. As seen, the model has a loitering phase followed by exponential expansion. The FLRW model will represent the cheese part of the Swiss-cheese model which is often referred to as the ``background" of the model.
\newline\newline
The FLRW model is transformed into an inhomogeneous LTB model by introducing an inhomogeneous big bang time. The LTB model is thus specified by having $k(r)$ and $M(r)$ equal to those of the FLRW model just described, and by $t_{bb}(r)$ given by
\begin{equation}
t_{bb}(r) = \frac{bb_{low_r}e^{bb_{sl}bb_s} + bb_{high_r}e^{bb_sr}}{e^{bb_{sl}bb_s} + e^{bb_sr}}.
\end{equation}
\begin{figure}
\centering
\includegraphics[scale = 0.6]{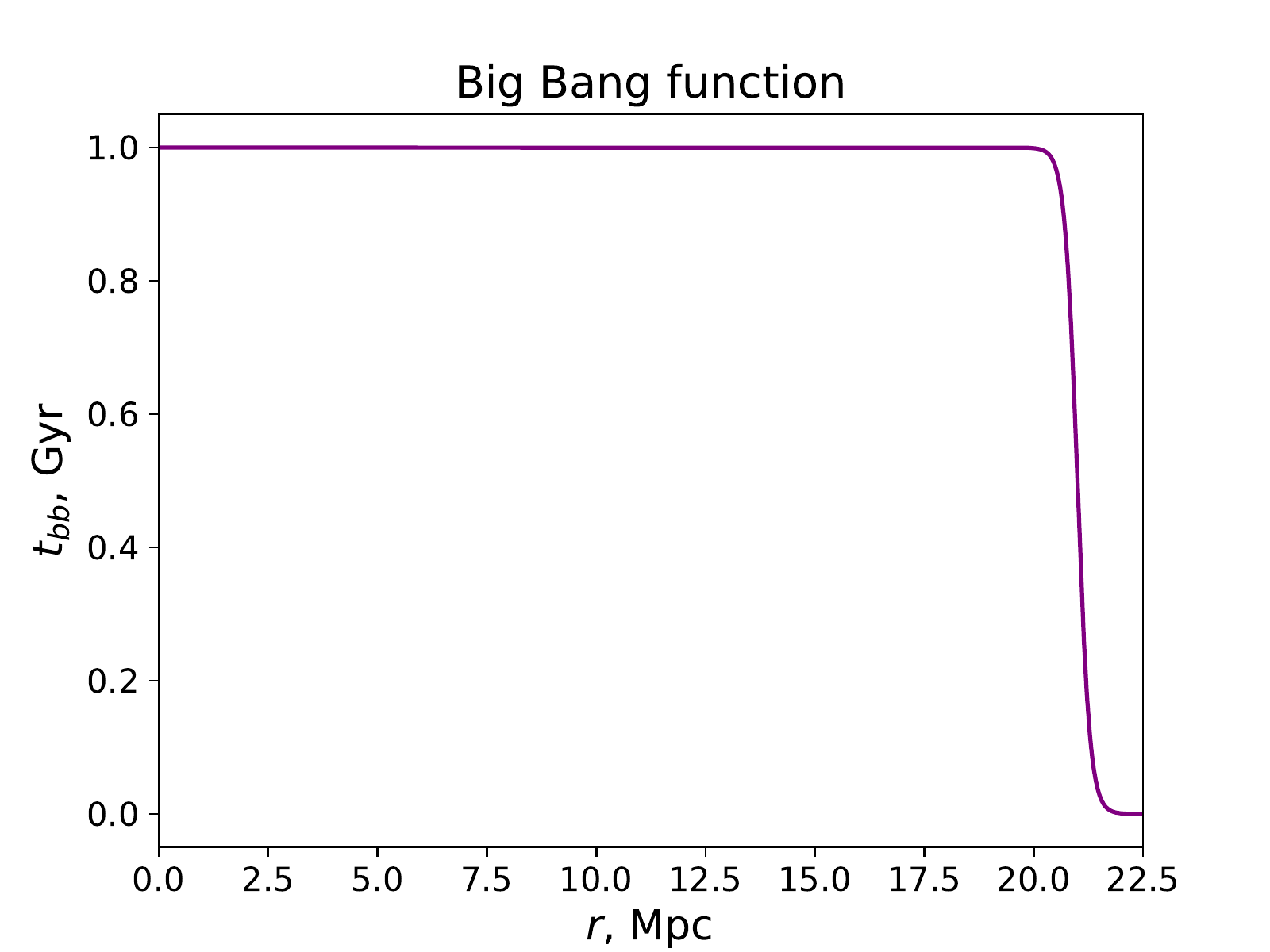}
\caption{Big bang time of the considered LTB model.}
\label{fig:tbb}
\end{figure}
The four parameters $bb_{low_r}, bb_{high_r}, bb_{s}$ and $bb_{sl}$ give, respectively, the value of $t_{bb}$ at the origin, the value of $t_{bb}$ at large $r$ values, the slope of the function and the slope location. For the model studied here, the values are chosen to be $1\text{Gyr}$,  $0\text{Gyr}$, $7\text{Mpc}^{-1}$ and $21\text{Mpc}$, respectively. The resulting $t_{bb}$ is shown in figure \ref{fig:tbb}. Clearly, the resulting LTB model will evolve as the FLRW model except near $r = bb_{sl} = 21\text{Mpc}$ and with the inner (low-r) part of the LTB model being $1$Gyr behind in its evolution compared to the outer ``cheese" part.
\newline\indent
Inhomogeneous big bang times are typically not considered in studies based on LTB models because an inhomogeneous big bang represents decaying modes \cite{decaying}. It is necessary to introduce an inhomogeneous $t_{bb}$ here simply because the pressure component from $\Lambda$ is homogeneous. In a more sophisticated model with spatial variation of the pressure component, the big bang time could be chosen to be homogeneous.
\newline\newline
\begin{figure}
\centering
\includegraphics[scale = 0.6]{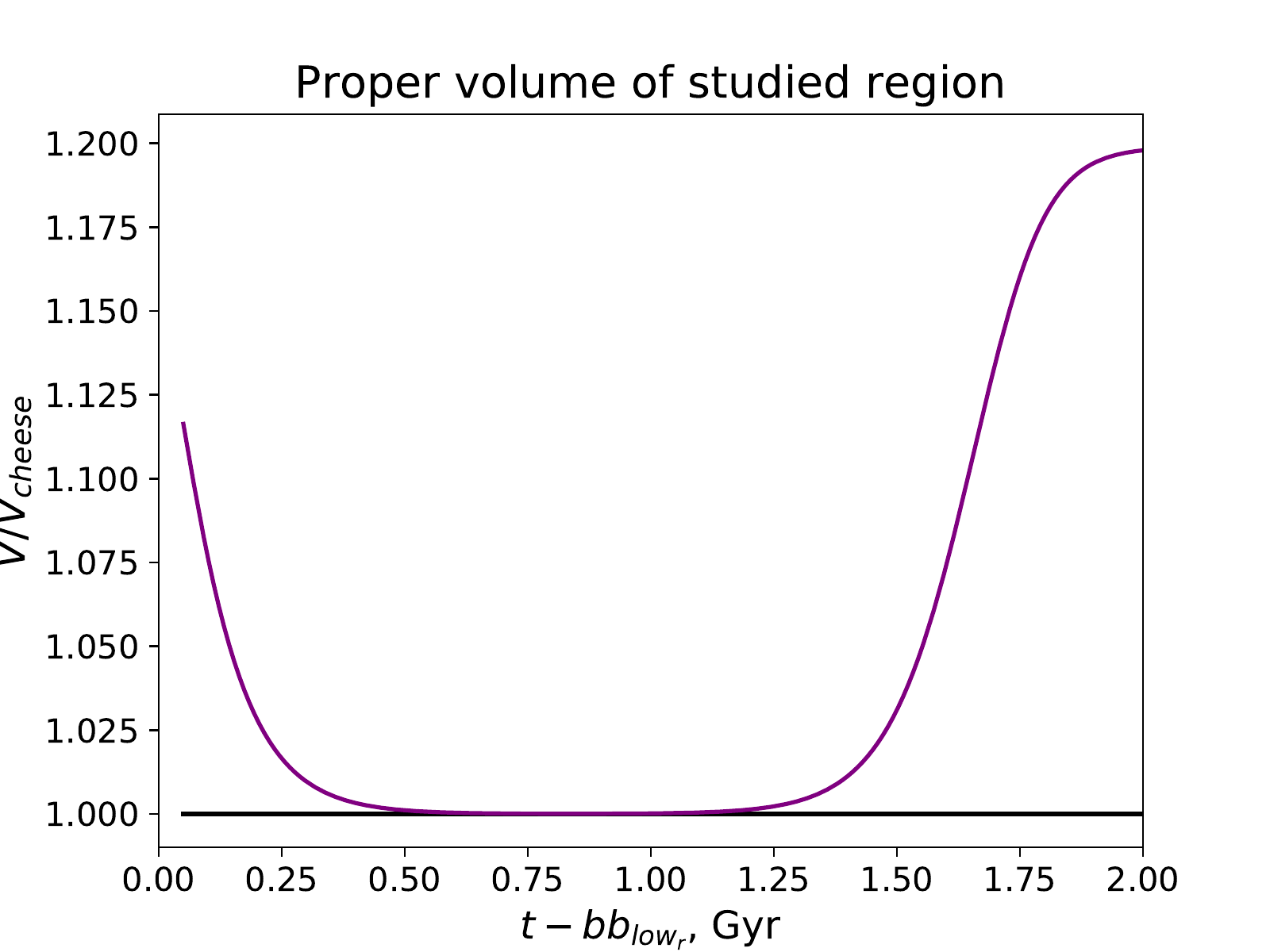}
\caption{Proper volume of spherical region compared to the proper volume of the FLRW cheese model. Volumes are computed over spherical regions centered at $r=0$ with $r\leq22.5$Mpc.}
\label{fig:volume}
\end{figure}

\begin{figure}
\centering
\includegraphics[scale = 0.6]{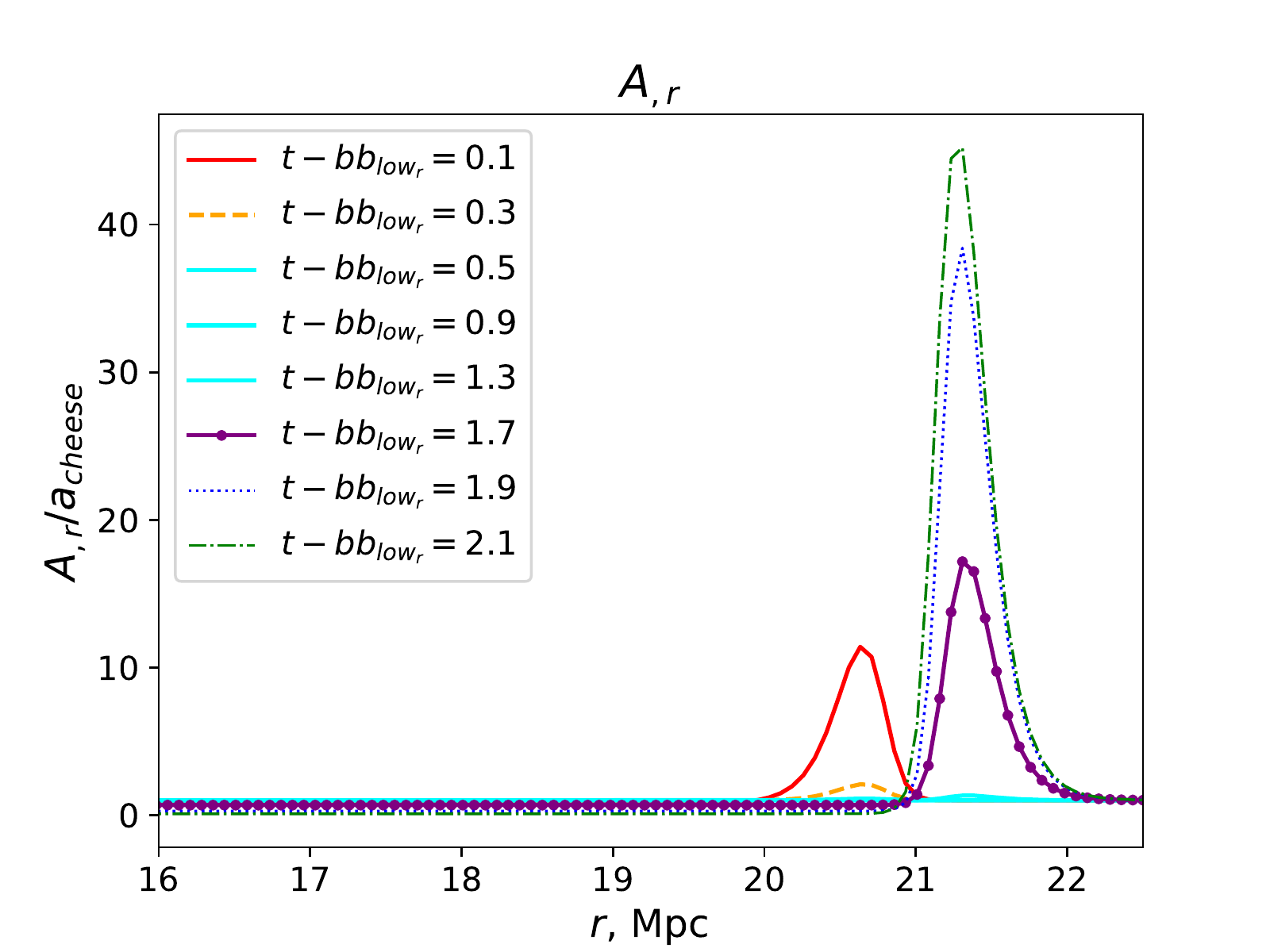}
\caption{$A_{,r}$, compared to the value of the scale factor of the cheese, $a_{cheese}$, at different times. At early and late times $A_{,r}$ is significantly larger than $a_{cheese}$ around the region where $\left( t_{bb}\right) _{,r}$ is large. At intermediate times, the entire LTB model is stuck in the loitering phase with $A\approx a_{cheese}r$ and therefore $A_{,r}/a_{cheese}\approx 1$ at these times. The lines representing these intermediate times are plotted with the same line type since they would be more or less indistinguishable anyway.}
\label{fig:Ar}
\end{figure}

\begin{figure}
\centering
\subfigure[]{
\includegraphics[scale = 0.6]{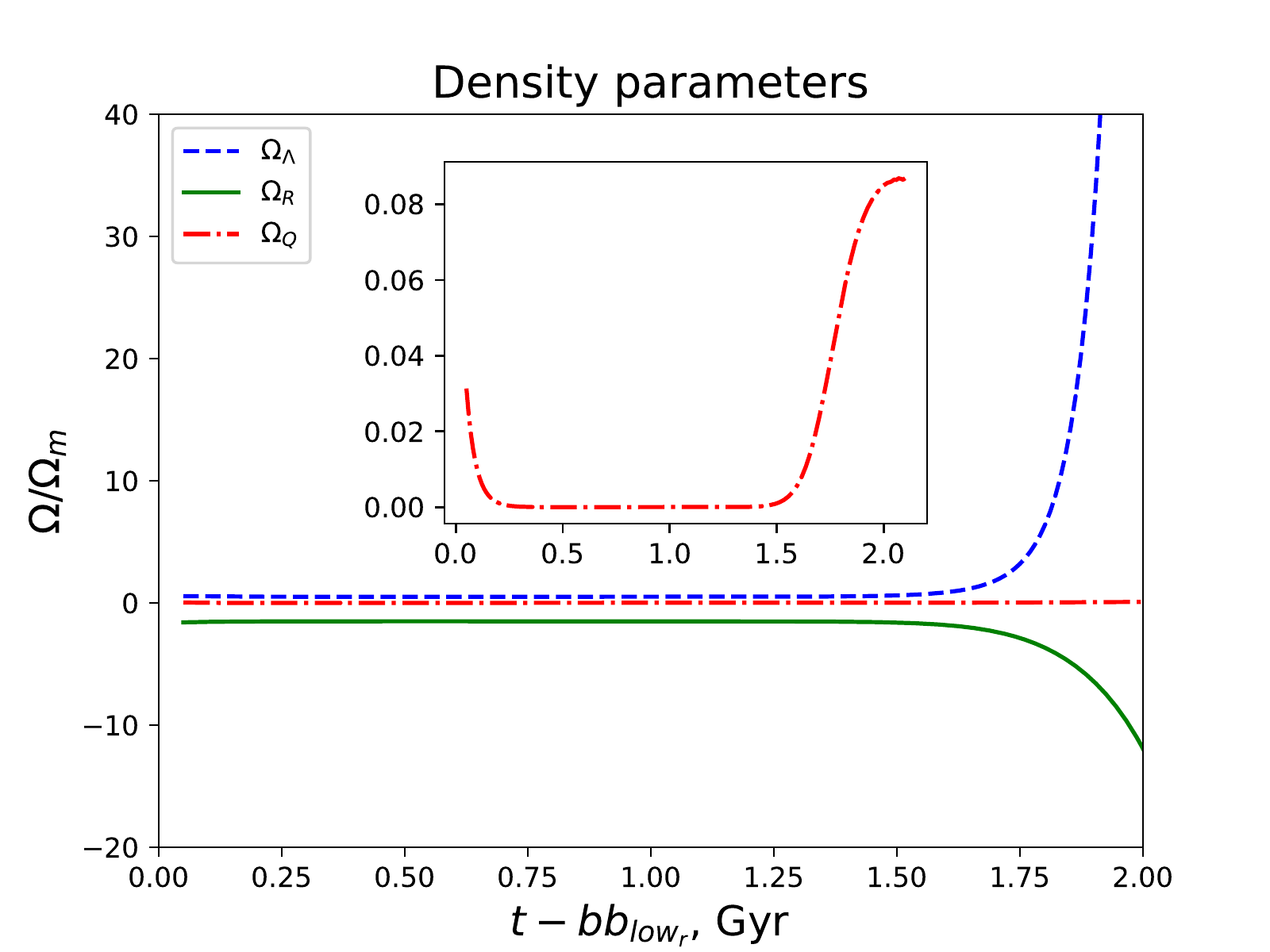}
}
\subfigure[]{
\includegraphics[scale = 0.6]{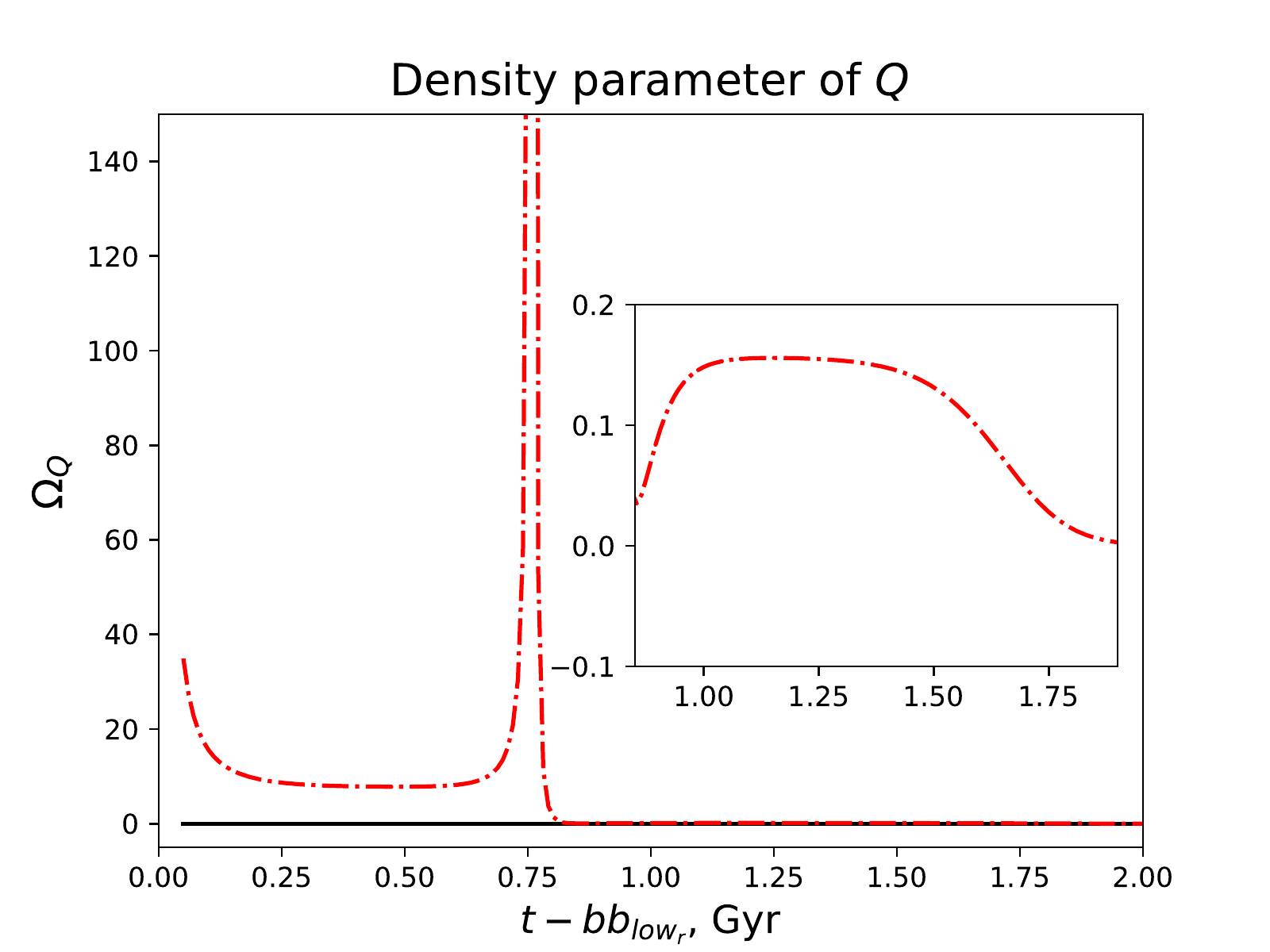}
}
\caption{Top: Average density parameters of the curvature, $\Lambda$ and $Q$ relative to the density parameter of dust. A close-up of the density parameter of $Q$ is included. Bottom: Density parameter of $Q$. Density parameters are defined as follows: $\Omega_m :=\frac{8\pi G_N\left\langle \rho\right\rangle}{3H_D^2}$, 
$\Omega_R:=-\frac{\left\langle ^{(3)}R\right\rangle}{6H_D^2}$, $\Omega_\Lambda:=\frac{\Lambda}{3H_D^2}$, and $\Omega_Q:=-\frac{Q_D}{6H_D^2}$. This is the usual way of defining density parameters in the Buchert average scheme and the density parameters defined this way add to 1.}
\label{fig:omegas}
\end{figure}
\begin{figure}
\centering
\includegraphics[scale = 0.6]{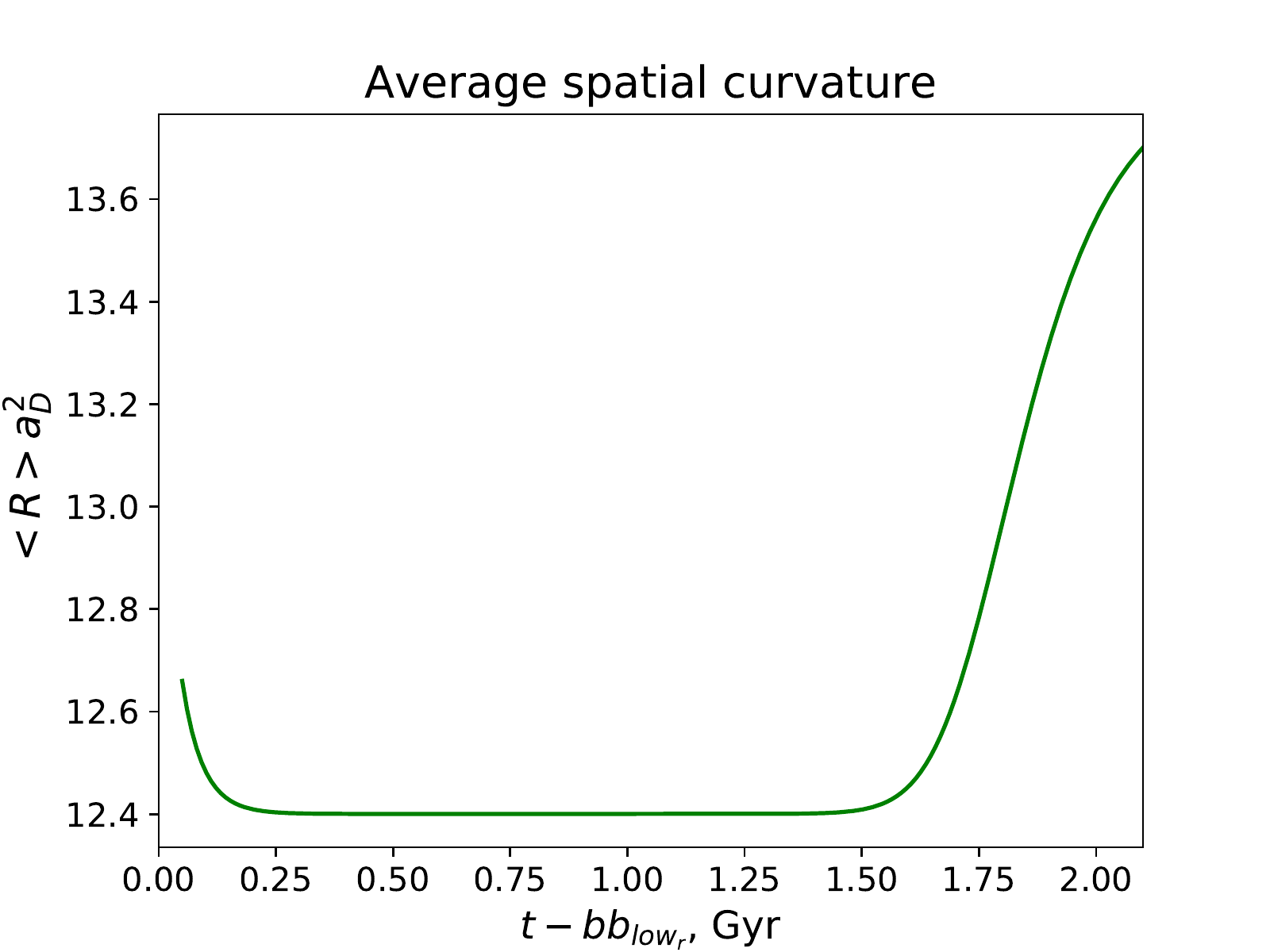}
\caption{Average spatial curvature multiplied by an un-normalized $a_D$.}
\label{fig:R}
\end{figure}

A fixed Swiss-cheese model can be constructed by distributing many copies of the LTB model in its cheese FLRW model. Since the cheese FLRW model used here has such large positive curvature, the fixed Swiss-cheese model should be constructed with periodic boundary conditions in accordance with the model's size. This could in principle be done in a manner similar to that in e.g. \cite{scSZ5} but there would be difficulties related to the background and average of the model not being Euclidean since a non-Euclidean space cannot be tiled by cubes. In addition, for a random distribution of single sized structures, the packing fraction of LTB models cannot go beyond approximately $0.64$ \cite{RCP_experimental}. Such small packing fraction would diminish the backreaction effect significantly. Therefore, the Swiss-cheese model will here be constructed on-the-fly by turning a light ray around when it reaches $r = 22.5$Mpc. This will not lead to a statistically homogeneous and isotropic spacetime, but if random impact parameters are chosen each time the light ray is turned back towards the structure, it will seem so from the light ray's point of view. The appropriate volume averages are based on the averaging domain\footnote{This domain size was chosen because $t_{bb}$ becomes smaller than $10^{-6}$ slightly before this. $10^{-6}$ is the order of precision used when solving the ODEs of section \ref{sec:exact_light} describing exact light propagation and can be considered the numerical precision used in this work.} $r\in [0,22.5]$. The resulting volume of the spherical region compared to the volume of the cheese FLRW model in the same r-interval is shown in figure \ref{fig:volume}. As seen, the volume of the LTB model is larger than the corresponding region in the cheese model, with a maximum of approximately $20\%$ in the studied time interval. It is somewhat unintuitive that the volume of the LTB model is {\em larger} than that of the cheese FLRW model as the cheese always has a larger scale factor than the inner part of the LTB region. The reason for the larger volume of the LTB model is seen in figure \ref{fig:Ar}. This figure shows the fraction $\frac{A_{,r}}{a_{cheese}}$ and shows that it becomes quite large in regions where $\left( t_{bb}\right) _{,r}$ is numerically large. Since $A_{,r}$ enters into the infinitesimal volume element ($dV = \frac{A_{,r}A^2}{\sqrt{1-k}}\sin^2(\theta)drd\theta d\phi$), its size is important for proper volumes. The reason $A_{,r}$ becomes so large is that it has to equal $a_{cheese}$ once at large values of $r$ where $t_{bb} \approx 0$. At times when the inner region of the LTB model has $A$ much smaller than $a_{cheese}r$, this means that $A_{,r}$ must be large near $r = bb_{sl}$. The inner region has $A$ much smaller than $a_{cheese}r$ at early and late times, but for an intermediate time period all spatial regions of the LTB model are in the loitering phase with approximately identical values of $A$.
\newline\indent
The evolution of the average density parameters are shown in figure \ref{fig:omegas} as fractions of the matter density parameter. The density parameter of the kinematical backreaction becomes numerically quite large but it only reaches a maximum of approximately $8.5\%$ of the average matter density parameter. As seen in figure \ref{fig:R}, this is large enough for the average spatial curvature, $\left\langle^{(3)}R \right\rangle $ to be visibly non-proportional to $a_D^{-2}$ during some time intervals.
\newline\newline

\begin{figure}
\centering
\subfigure[]{
\includegraphics[scale = 0.6]{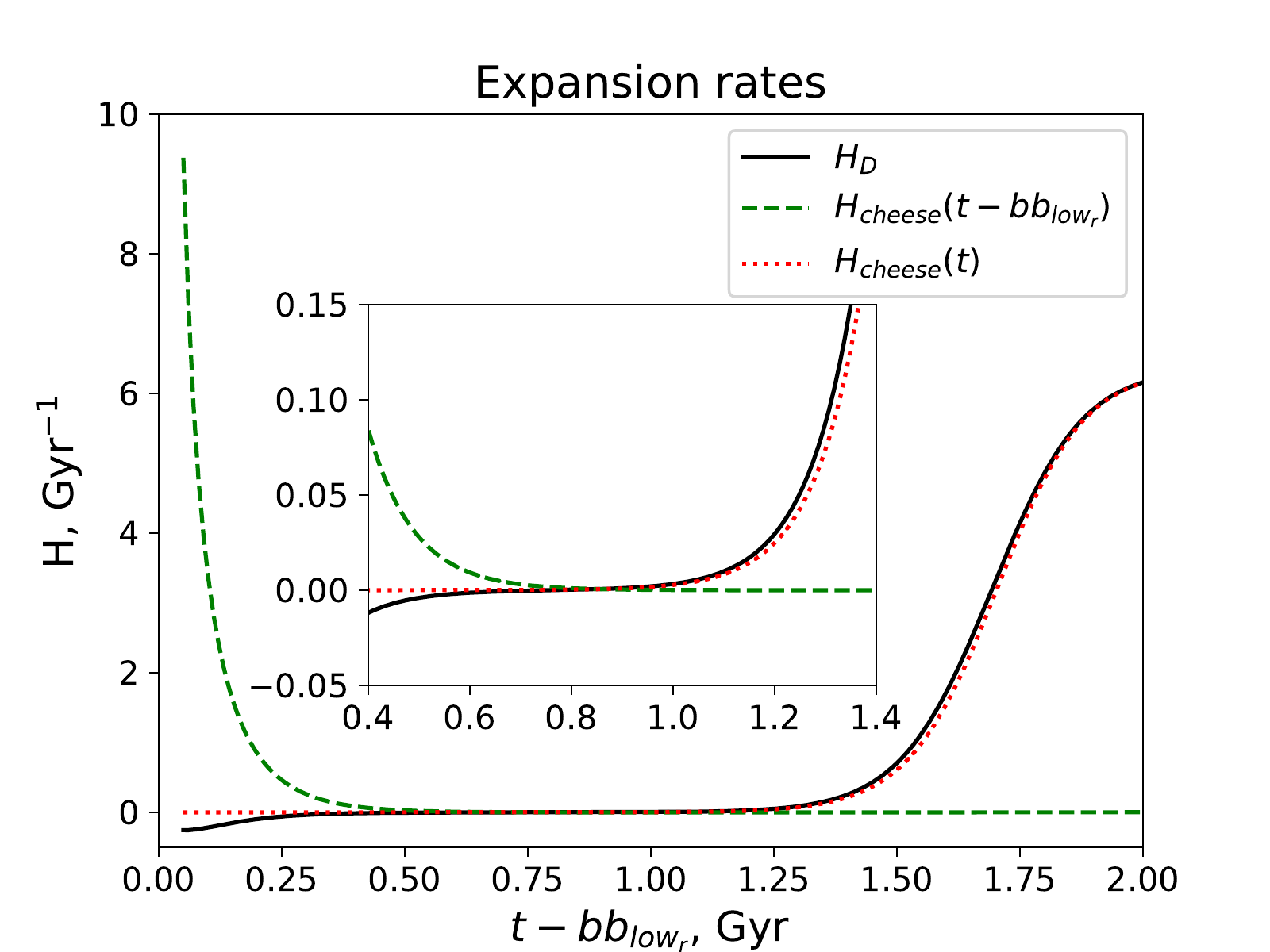}
}
\subfigure[]{
\includegraphics[scale = 0.6]{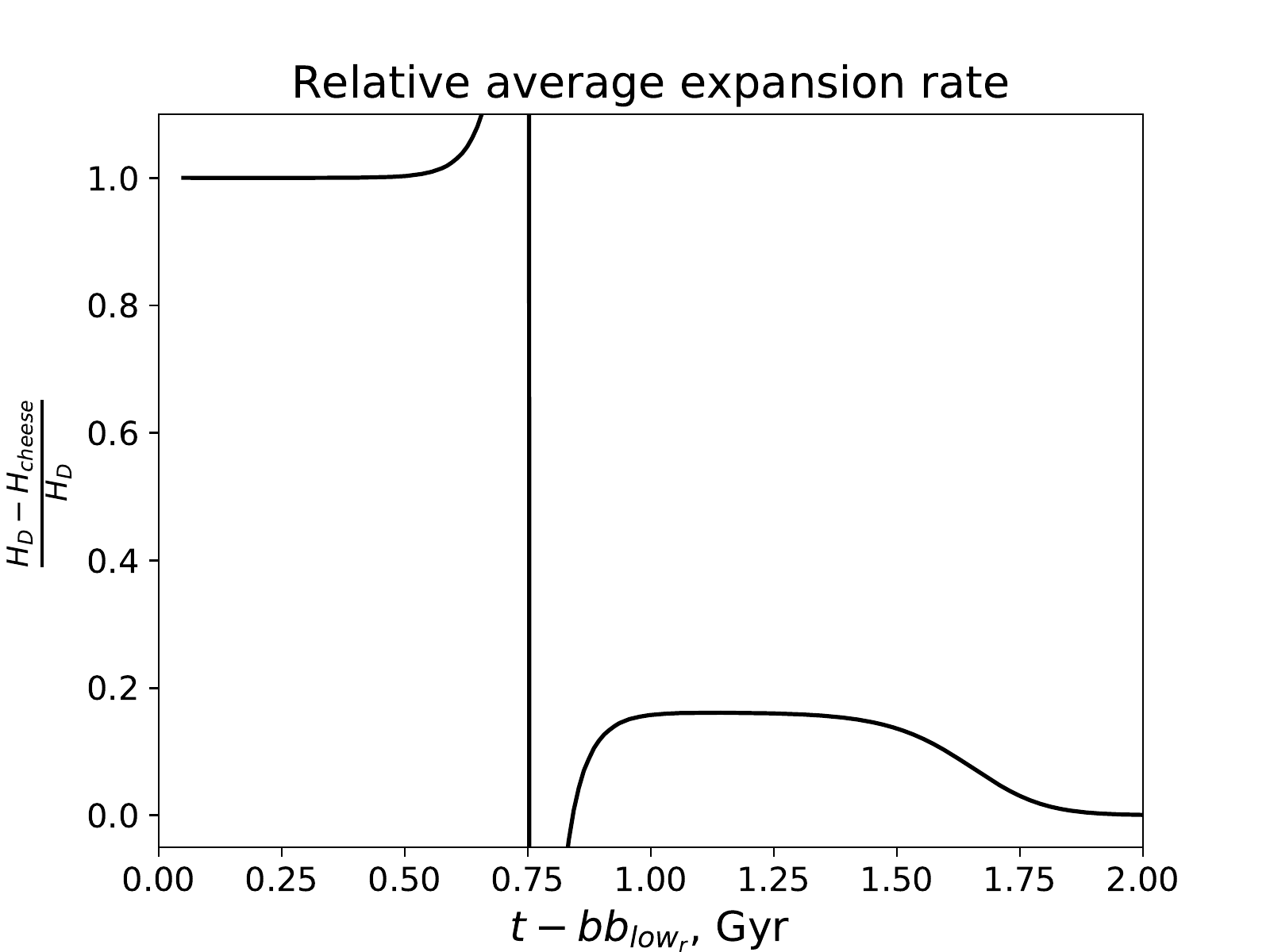}
}
\caption{Top: Average Hubble parameter compared to the Hubble parameter of the FLRW cheese. The latter is shown both at $H_{cheese}(t)$ and $H_{cheese}(t-bb_{low_r})$ with the latter corresponding to the expansion rate of the LTB region near $r = 0$. Bottom: Average Hubble parameter relative to the Hubble parameter of the cheese FLRW model. There is a divergence ar $t-bb_{low_r}\approx 0.75$ where $H_D$ crosses 0.}
\label{fig:H}
\end{figure}
The average expansion rate of the region is shown in figure \ref{fig:H}. At early times, the cheese is in the loitering phase while $H_D$ is clearly non-zero. In fact, $H_D$ is negative even though the cheese FLRW model always has a non-negative expansion rate. The negative contribution to $H_D$ comes from the region where $t_{bb,r}$ is large. $H_D$ and $H_{cheese}$ become hard to distinguish from each other for $t-bb_{low_r}\gtrsim  0.5$Gyr. However, a plot of $\frac{H_D-H_{cheese}}{H_D}$ (figure \ref{fig:H}) shows that for $t-bb_{low_r}\in [0.8,1.5]$, the average Hubble parameter and the Hubble parameter of the cheese deviate from each other by approximately $16\%$. From $t-bb_{low_r}\approx 2$Gyr and onward, the cheese begins to dominate the averages and backreaction becomes negligible.
\newline\newline
Expressions for average quantities of the LTB model are given in appendix \ref{app:averages}.
\newline\indent
The following two sections review light propagation in LTB models and based on volume averages, and section \ref{sec:results} shows results from applying these formalisms to the example model described here.

\section{Exact light propagation}\label{sec:exact_light}
Although the global curvature of the model studied here is quite large, the geometric optics approximation as well as the geodesic deviation equation should both still be valid. Thus, the path of a light ray is given by the null geodesic equation. For the LTB model, the null geodesic equation leads to the following four equations for the components of the null tangent vector $k^{\mu}$ (dots denote differentiation with respect to the affine parameter $\lambda$):
\begin{equation}
\dot k^t = \frac{1}{2}\left( R_{,t}\left( k^r\right) ^2 + F_{,t}\left( k^{\theta}\right) ^2 + P_{,t}\left( k^{\phi}\right) ^2 \right) 
\end{equation}
\begin{equation}
\dot k^r = \frac{1}{2R}\left( -2\dot R k^r + R_{,r}\left( k^r\right) ^2 + F_{,r}\left( k^{\theta}\right) ^2 + P_{,r}\left( k^{\phi}\right) ^2 \right) 
\end{equation}
\begin{equation}
\dot k^{\theta} = \frac{1}{2F}\left( -2\dot F k^{\theta} + P_{,\theta}\left( k^\phi\right) ^2 \right) 
\end{equation}
\begin{equation}
\dot k^{\phi} = \frac{\dot P}{P}k^{\phi}
\end{equation}
The above correspond to the LTB line element written as $ds^2 = -dt^2 + Rdr^2 + Fd\theta^2 + Pd\phi^2$.
\newline\newline
The angular diameter distance, $D_A$, can be computed from the geodesic deviation equation. As shown in e.g. \cite{arbitrary_spacetime}, the geodesic deviation equation can be rewritten to yield the transport equation
\begin{equation}\label{D_dot}
\ddot D^a_b = T^a_cD^c_b.
\end{equation}
If initial conditions are chosen such that $k^t_{initial} = -1$, the square root of the absolute value of the determinant of $D$ is the angular diameter distance along the light ray. The tidal matrix is given by
\begin{equation}
 T_{ab} = 
  \begin{pmatrix} \mathbf{R}- Re(\mathbf{F}) & Im(\mathbf{F}) \\ Im(\mathbf{F}) & \mathbf{R}+ Re(\mathbf{F})  \end{pmatrix} ,
\end{equation}
where $\mathbf{R}: = -\frac{1}{2}R_{\mu\nu}k^{\mu}k^{\nu} = -4\pi G_N\rho \left( k^t\right) ^2$ and $\mathbf{F}:=-\frac{1}{2}R_{\alpha\beta\mu\nu}(\epsilon)^{\alpha}k^{\beta}(\epsilon)^{\mu}k^{\nu}$. Here, $R_{\mu\nu}$ denotes the Ricci tensor, $R_{\alpha\beta\mu\nu}$ the Riemann tensor and $\epsilon^{\mu} := E_1^{\mu} - iE_2^{\mu}$ with $E_1^{\mu}, E_2^{\mu}$ the screen space basis vectors (screen space is the 2D Euclidean space orthogonal to the light ray direction as seen from the observer's rest frame). In order to solve the tidal equation along a light path, $E_1^{\mu}$ and $E_2^{\mu}$ must be parallel transported along the light ray. Their initial conditions must be set so that they are orthonormal and orthogonal to the null tangent vector. This is fulfilled if the initial conditions are set according to
\begin{equation}
\left( E_1^{\mu}\right)_0 \propto \left(0, -\frac{1}{k^rR}\left(\frac{(k^{\theta})^2}{k^{\phi}}F + k^{\phi}P\right), \frac{k^{\theta}}{k^{\phi}} ,1 \right) 
\end{equation}
\begin{equation}
\left( E_2^{\mu}\right)_0 \propto \left(0,0,-\frac{k^{\phi}}{k^{\theta}}\sin^2(\theta),1 \right) .
\end{equation}
If initial conditions are set such that either of $k^r,k^{\theta},k^{\phi}$ is zero, the above initial conditions clearly do not apply. For radial initial conditions one can instead choose
\begin{equation}
\left( E_1^{\mu}\right) _0 = \left(0,0,\frac{1}{\sqrt{R}},0 \right) 
\end{equation}
\begin{equation}
\left( E_2^{\mu}\right) _0 = \left(0,0,0,\frac{1}{\sqrt{P}} \right).
\end{equation}
By simultaneously solving the null geodesic equations, the transport equation and the equations for parallel transport of the screen space basis vectors along the light path, the redshift-distance relation along the light ray bundle can be obtained. These exact light propagation results can then be compared to results obtained from formalisms based on volume averaged quantities in order to see how accurate such formalisms are in reproducing observations. The volume average based formalisms considered here are described in the next section (after a small interlude in the following subsection). The Riemann components and $\mathbf{F}$ for the LTB metric are shown in appendix \ref{app:light}.

\subsection{Light propagation in closed FLRW models}
Since the studied universe model is quite small, some remarks regarding peculiarities of $D_A$ in closed FLRW models are in order.
\newline\newline
As is well known, the angular diameter distance as a function of redshift may in a closed FLRW model be written as
\begin{equation}
D_A(z) = \frac{a_0}{1+z}R_0\sin\left( \frac{R_0}{a_0}\int_{0}^{z}\frac{dz'}{H(z')}\right), 
\end{equation}
where $R_0$ is the curvature radius.
\newline\newline
The expression shows that for increasing redshifts, $D_A$ will grow initially but then it will decrease until it reaches zero. This happens when the light ray reaches the antipodal point of the Universe compared to the observer. After this point, the image of a light bundle is flipped which is represented by $D_A$ becoming negative. This flip is not captured by the formalism described in the previous section; $D_A$ was there noted to be given as the square root of the absolute value of the determinant of $D$ and is hence always non-negative. Therefore, to facilitate the comparison of $D_A$ computed using different schemes, all angular diameter distances presented in section \ref{sec:results} will be shown as absolute values.

\section{Average light propagation}\label{sec:av_light}
As mentioned in the introduction, at least two different schemes for relating spatial averages to observations within the Buchert averaging scheme have been proposed. In one scheme (\cite{av_obs1,av_obs2}), the mean/average of the observed redshift and angular diameter distance are related to the average density, the volume scale factor and the average expansion rate by invoking considerations regarding spatial statistical homogeneity and isotropy as well as the mean characteristic evolution time of large scale structures. This scheme will in the following be referred to as the covariant scheme. In the other approach (\cite{template1,template2}), a ``template" metric is introduced which resembles the FLRW metric but which uses the volume-averaged scale factor and the spatially averaged curvature. This scheme will in the following be referred to as the template scheme.
\newline\newline
The covariant and template schemes for describing average light propagation will be summarized in the two following sections. The two schemes will then be compared in section \ref{subsec:Q_size}.

\subsection{The covariant scheme}\label{subsec:covariant}
As described in the introduction, the spacetime setup used in the Buchert averaging formalism for a dust universe considers a foliation of spacetime with space orthogonal to the dust velocity field $u^{\mu}$. In such a frame, a fundamental observer comoving with the dust sees light redshifted according to the relation 
\begin{equation}
\frac{d\gamma}{\gamma}= \left( \frac{1}{3}\Theta + \sigma_{\mu\nu}e^{\mu}e^{\nu}\right) ,
\end{equation}
where $\gamma$ is the wavelength and $e^{\mu}$ is a unit vector proportional to the spatial direction of the null geodesic tangent vector, i.e. $e^{\mu} = \frac{u^{\mu}-k^{\mu}}{u^{\nu}k_{\nu}}$ (see e.g. \cite{god_bog,Ehlers}).
\newline\indent
In \cite{av_obs1} (see also \cite{av_obs2} for general perfect fluids), this relationship is integrated to obtain the result
\begin{equation}\label{eq:z_exact}
1+z = e^{\int_{t_e}^{t_0}dt\left(\frac{1}{3}\Theta + \sigma_{\alpha}^{\beta}e^{\alpha}e_{\beta} \right)  } ,
\end{equation}
where the subscript $e$ denotes evaluation at the point of emission.
\newline\newline
In \cite{av_obs1} it is then argued that if space is statistically homogeneous and isotropic with structures evolving slowly compared to the time it takes for a light ray to traverse an assumed homogeneity scale, then the observed redshift is given by
\begin{equation}
1+z^C: = e^{\frac{1}{3}\int_{t_e}^{t_0}dt\left\langle \Theta\right\rangle}= \frac{1}{a_D} ,
\end{equation}
with deviations due to statistical fluctuations, assuming that $D$ is larger than the homogeneity scale. The superscript $C$ on the redshift is used to indicate that the expression describes the redshift according to the ``covariant" scheme.
\newline\indent
The assumptions listed above also lead to a differential equation approximately describing the angular diameter distance up to statistical fluctuations: 
\begin{equation}\label{eq:DA_covariant}
H_D\frac{d}{dz^C}\left(\left(1+z^C \right) ^2H_D \frac{dD_A^C}{dz^C}\right) = -4\pi G\left\langle \rho\right\rangle D_A^C
\end{equation}
Thus, assuming that statistical fluctuations become negligible upon averaging over a given data set, the redshift-distance relation is given by the set $\left(D_A^C,z^C \right) $, where $z^C$ is related to the volume scale factor exactly as the redshift in the FLRW model, but with $D_A^C$ not in general being given as in the FLRW models.
\newline\indent
The average redshift $z^C$ is not monotonic for the example model of section \ref{subsec:example} so the differential equation for $D_A^C$ must in practice be written and solved for in terms of $t$ instead of $z$. Following the considerations in \cite{av_obs1}, the appropriate differential equation is seen to be
\begin{equation}
\frac{d^2D_A^C}{dt^2} = -4\pi G_N\left\langle \rho\right\rangle D_A^C + H_D\frac{dD_A^C}{dt}
\end{equation}
The equation must be solved with the initial conditions $\frac{dD_A}{dt} = -1, D_A = 0$.

\subsection{The template scheme}\label{subsec:template}
In \cite{template1} (see also e.g. \cite{template2,template3}) a different on-average description of the redshift-distance relation is obtained by introducing a template metric. Specifically, it is in \cite{template1} noted that while the Buchert equations describe the average kinematical qualities of a flow orthogonal spacetime foliation, a distance measure also needs to be introduced. It is then noted that Ricci flow renormalization \cite{Ricci_flow} constitutes a homogenization procedure for inhomogeneous spatial hypersurfaces of cosmological models which produces a constant curvature space at any given time. This curvature may be different at different times. For a given choice of averaging domain, each hypersurface is therefore characterized by the volume scale factor as well as a time/hypersurface-dependent curvature parameter, $k_D(t)$.
\newline\newline
Using the above considerations, light paths are within the template setting postulated (on-average) to follow the relation
\begin{equation}\label{eq:drdt}
\frac{dr}{dt}=\frac{1}{a_D}\sqrt{1-k_D(t)r^2}.
\end{equation}
A template metric is then introduced by requiring that light paths are null according to the template metric and hence the template metric is given by
\footnote{Note that this metric combined with the average dust component is not required or expected to fulfill the Einstein equation. Indeed, it has been shown (\cite{syksy_kt}) that the metric does not correspond to a spatially homogeneous universe when inserted into the Einstein equation. See however \cite{Stichel} regarding its relation to solutions to the Einstein equation.}
\begin{equation}\label{eq:template_metric}
ds^2_D=-dt^2 + a_D^2\left(\frac{dr^2}{1-k_D(t)r^2} + d\Omega^2 \right).
\end{equation}
It is in \cite{template1} suggested that the curvature function $k_D(t)$ should be related to the average intrinsic curvature according to $k_D(t)=\frac{1}{6}a_D^2\left\langle ^{(3)}R\right\rangle $.
\newline\newline
With this template metric at hand for defining average distances, the angular diameter distance is given by $D^T_A:=r_D(z^T)a_D(z^T)$, where the superscript $T$ is introduced on the effective angular diameter distance and redshift to indicate that they are based on the ``template" scheme. $r_D(z^T)$ can be found by solving equation \eqref{eq:drdt}.
\newline\indent
According to the definition $1+z:=\frac{\left( k_{\mu}u^{\mu}\right) _e}{\left(k^{\nu}u_{\nu} \right) _O}$, the template metric leads to an effective redshift given by
\begin{equation}
z^T:=\frac{1}{a_D}e^{\frac{1}{2}\int_{t_e}^{t_0}\frac{r^2k_{D,t}}{1-k_Dr^2}dt}-1.
\end{equation}
The redshift can alternatively be obtained by simultaneously solving equation \eqref{eq:drdt} and the equation
\begin{equation}\label{eq:dkdt}
\frac{dk^t}{dt}=\frac{1}{2}k^t\left(2H_D - \frac{k_{D,t}r^2}{1-k_Dr^2} \right).
\end{equation}
The above is simply the geodesic equation for the time-component of the null geodesic tangent vector, $k^t$, according to the template metric. The redshift can then be obtained as $1+z^T = \frac{\left(u^{\alpha}k_{\alpha} \right)_e }{\left(u^{\beta}k_{\beta} \right) _0} = \frac{k^t_e}{k^t_0}$. Within the template scheme, the relationship between redshift and distance is thus on-average postulated to be described by the pair $\left(D_A^T,z^T \right) $ where both $D_A^T$ and $z^T$ will in general deviate from FLRW versions.
\newline\newline
The FLRW line element corresponding to equation \eqref{eq:template_metric} can be re-written to the well-known form $ds^2=-dt^2+a^2\left( dw^2+R_0^2\sin^2\left(\frac{w}{R_0} \right)d\Omega^2\right) $ with $R_0$ the curvature radius. Because of the time dependence of $k_D$, a similar coordinate transformation, $r\rightarrow w$ for $r = \frac{1}{\sqrt{k_D}}\sin\left(w\sqrt{k_D} \right) $, of the line element in equation \eqref{eq:template_metric} would not simply lead to the line element $ds^2=-dt^2+a_D^2\left(dw^2+\frac{1}{k_D}\sin^2\left(w\sqrt{k_D} \right)d\Omega^2\right) $. However, the template metric is not meant to represent a solution to the Einstein equation but it rather chosen based specifically on light paths. Therefore, the template scheme should be considered equally valid with a template metric written in line with the regular FLRW-type metric for $w$, i.e. the template metric may be written as
\begin{equation}\label{eq:dw}
ds^2=-dt^2+a_D^2\left( dw^2+\frac{1}{k_D}\sin^2\left(w\sqrt{k_D} \right)d\Omega^2\right).
\end{equation}
This version of the template metric is used in e.g. \cite{virialisation}.
\newline\indent
The template metric in equation \eqref{eq:dw} is important here as the goal is to study light rays circumnavigating the model universe several times. This is not possible using the template metric written as in equation \eqref{eq:template_metric} since division with zero occurs when $r = \frac{1}{\sqrt{k_D}}$. With the template metric in the form of equation \eqref{eq:dw}, the equations to be solved are instead
\begin{equation}\label{eq:DAz_dw}
\begin{split}
\frac{dw}{dt} = -\frac{1}{a_D}\\
\frac{dk^t}{dt} = H_Dk^t\\
D_A^T = \frac{a_D}{\sqrt{k_D}}\sin\left(w \sqrt{k_D}\right) .
\end{split}
\end{equation}
These equations are solved with the initial conditions $w = 0$ and $k^t = -1$.
\newline\newline
The redshift and angular diameter distance given by equation \eqref{eq:DAz_dw} and by equations \eqref{eq:drdt} and \eqref{eq:dkdt} are not generally the same. This is clearly an issue for the template scheme in its current form as there is no apparent theoretical justification for using one version over the other. Since the latter system simply cannot be used here, equation \eqref{eq:DAz_dw} will be used and referred to at the template scheme.

\subsection{Comparison of the two schemes}\label{subsec:Q_size}
In general, the covariant and template schemes do not predict identical redshift-distance relations. However, as mentioned in the introduction, an average evolution of the Universe may be related to a small kinematical backreaction that drives the average evolution to slowly evolve between different FLRW solutions such that $a_D^2\left\langle^{(3)}R \right\rangle _D\approx$ const.. In such cases, the covariant and the template schemes yield the same description for the average redshift distance relation\footnote{Both versions of the template scheme discussed in the previous subsection reduce to this relation in the given limit. The version of the template scheme used here leads to the relation for any size of $Q$.}:
\begin{equation}
z^T = z^C = \frac{1}{a_D}-1=:z^D
\end{equation}
\begin{equation}
\begin{split}
D_A^T=D_A^C=\frac{1}{\left( 1+z_D\right)\sqrt{K}}\sin\left(\sqrt{K}\int_{0}^{z^D}\frac{d\hat z^D}{H_D} \right),
\end{split}
\end{equation} 
with $K:=\frac{1}{6}\left\langle^{(3)}R \right\rangle_{|t=t_0}$.
\newline\newline
Since the redshift-distance relation in this case takes the same form as in the FLRW models, a universe with this type of average dynamics will not fail FLRW geometry consistency relations such as those proposed in \cite{copernican,parallax,sum_rule}. This situation therefore illustrates the general point (also noted in e.g. \cite{parallax}) that observational verifications of FLRW geometry tests need not imply that the Universe is well described by a fixed FLRW model. However, while these expressions look exactly as in the FLRW case, the non-vanishing of $Q$ implies that $a_D$ can evolve very differently than in a fixed FLRW spacetime and hence very different redshift-distance relations can result.

\section{Light propagation results}\label{sec:results}
This section presents the results obtained from propagating individual light rays through the example model of section \ref{subsec:example}. All results are for an observer placed a $t = t_0$ with $t_0 -bb_{low_r}= 1.8$Gyr. For this value of $t_0$, the expansion rate of the background has not yet become exponential and its scale factor is still fairly small. A large scale factor and/or expansion rate will lead structures to evolve fast compared to the time it takes a light ray to traverse the homogeneity scale of the model. When this happens, it is no longer reasonable to expect that the covariant or template scheme can describe the redshift-distance relation well on average; when a structure evolves significantly while a light ray traverses it, cancellations between local fluctuations from the average in over- and underdense regions can no longer be expected. Similarly, some caution should be taken when interpreting the high-z part of the redshift-distance relations presented in the figures of this section since the low-r part of the LTB structure has a large $H$ at small $t$. Since the effect is cumulative, this is less important than large $a$ or $H$ at times closer to $t_0$.
\newline\indent
The results shown in this section are for ``radial" light rays i.e. light rays that have been propagated radially through consecutive LTB structures. This is to facilitate a simpler presentation of the results and results obtained by considering light rays with random impact parameters at turnaround points (i.e. when $r = 22.5$Mpc is reached) are commented on where appropriate.
\newline\newline
\begin{figure*}
\centering
\subfigure[]{
\includegraphics[scale = 0.48]{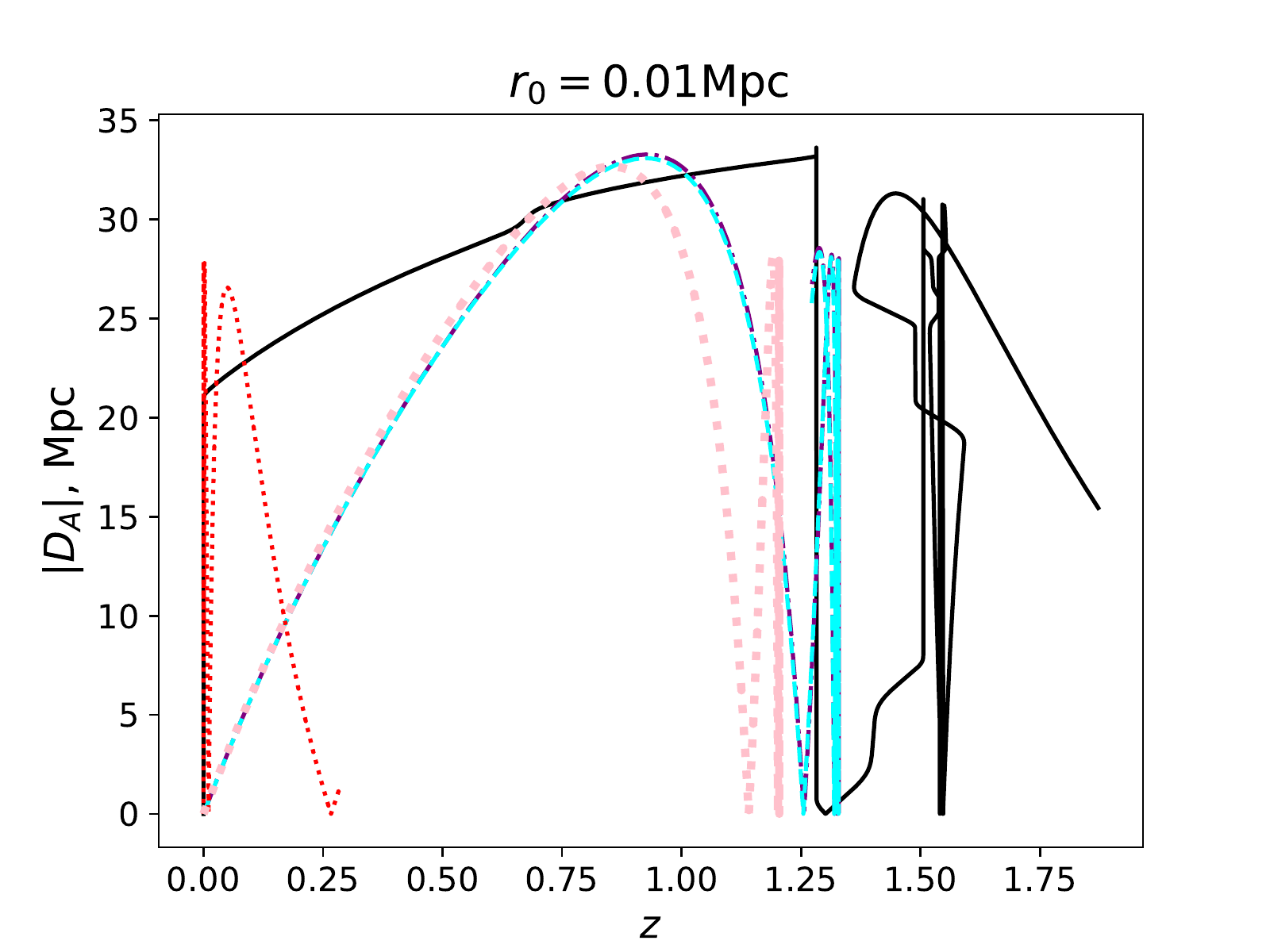}
\label{subfig:a}
}
\subfigure[]{
\includegraphics[scale = 0.48]{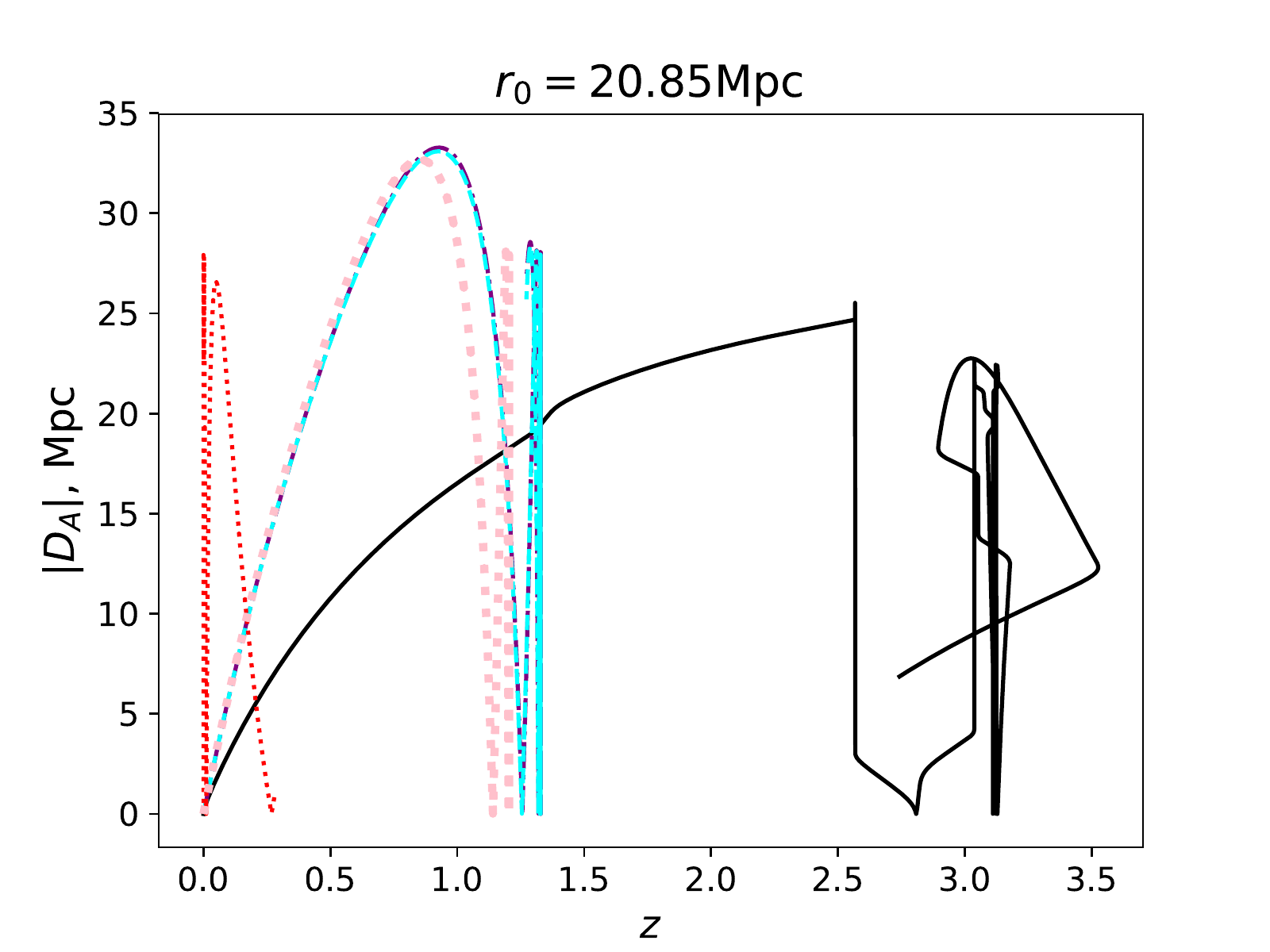}
}
\subfigure[]{
\includegraphics[scale = 0.6]{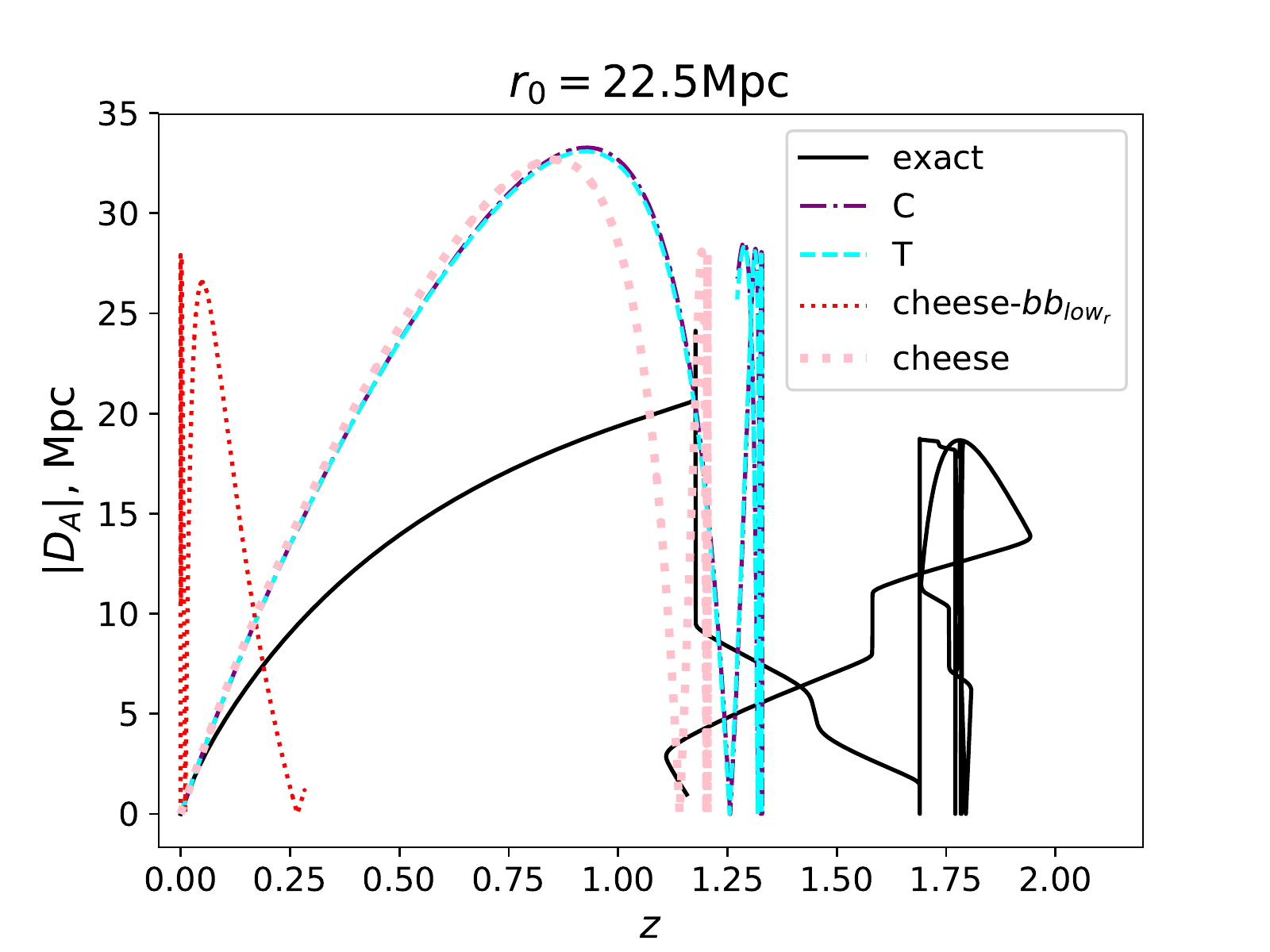}
}
\caption{Redshift-distance relation along radial light rays with the observer placed at different radial coordinates (indicated at the top of each figure by the value of $r_0$). Labels given in the bottom figure are valid for all three plots. The lines labeled ``exact" represent the redshift-distance relations obtained by propagating light rays using the exact LTB metric of the Swiss-cheese spacetime. The lines labeled ``C" and ``T" denote the predictions of the covariant and template schemes while ``cheese" denotes the redshift-distance relation of the cheese and ``cheese-$bb_{low_r}$" indicates the same but for a model shifted by $t = bb_{low_r}$ such that it corresponds to the central part of the LTB model.}
\label{fig:DA}
\end{figure*}
Figure \ref{fig:DA} shows the redshift-distance relation obtained by propagating a light ray radially back and forth in the interval $r\in [0,22.5]$Mpc. The redshift-distance relation is compared with those predicted by the covariant and template schemes as well as the redshift-distance relation in homogeneous FLRW models corresponding to the cheese and the central part of the LTB model.
\newline\indent
The first point to note regarding figure \ref{fig:DA} is the somewhat chaotic behavior of the exact redshift-distance relations. This is due to the significant differences in the local expansion rates in the inner parts of the LTB model and the cheese. Take for instance the exact redshift-distance relation in figure \ref{subfig:a}. The redshift-distance relation is initially nearly vertical because the observer is placed in the central part of the LTB model at a time where that region is in a loitering phase. While the light ray propagates towards the cheese it is barely redshifted at all while the corresponding angular diameter distance, however, grows. The light ray then moves into a region with quite fast local expansion rate so that its redshift-distance relation becomes more flat and then the light ray moves back into a loitering region. The light ray keeps moving between loitering and non-loitering regions but at earlier times it is the cheese that loiters. At these times, the inner part of the LTB region is expanding and has a dust density much larger than the cheese which is why the local blueshifting appears in the graphs. Specifically, blueshifting easily appears in inhomogeneous models such as Szekeres models and LTB models where it can, for instance, appear when a light ray propagates in a direction of decreasing density. This can for instance be seen clearly in figures 7-9 in \cite{dig1} and the phenomenon was studied in \cite{blueshift}.
\newline\indent
It is clear from the figure that neither the covariant nor the template scheme does a good job in reproducing the redshift-distance relation along individual light rays. This is not too surprising as the schemes are only expected to be good ``on average" i.e. after averaging over many light rays. This requires propagating many light rays with random impact parameters at each turnaround point - for a single observer. However, by studying a couple of such light rays it becomes clear that this only has a slight impact on the results. It is therefore much more interesting to show results for observers placed at different values of the radial coordinate as the observer position has a big impact on the results. In figure \ref{fig:DA}, results are shown for an observer close to the origin, an observer placed at the ``edge" of the spherical region under consideration, and an observer at a position where the local Hubble parameter is approximately equal to the average Hubble parameter at observation time. The three light rays clearly have very different redshift-distance relations. By comparing the $D_A$ amplitudes of the individual redshift-distance relations in figure \ref{fig:DA}, it may be noticed that the $D_A$ amplitude of the exact light rays appear to be larger than that ``predicted" by the template and covariant approximation schemes if the observer is placed in the central part of the LTB region, but smaller if the observer is placed near or in the cheese. Averaging over many observer positions might therefore lead to a better agreement between the approximate and exact results. Since such an averaging cannot be made with real observations, it is not generally appropriate for theoretical studies but it may be reasonable if models have large local effects (such as here) that are not expected to exist in the real universe.
\newline\indent
While figure \ref{fig:DA} shows a clear difference between the exact and approximate redshift-distance relations, the difference between the predictions of the covariant scheme and the template scheme is not distinguishable in the figure. There is a small difference though, with the covariant scheme having slightly larger amplitude of $D_A$ than the template scheme. Since the agreement between the exact and approximate results is so poor, it is not possible to conclude that either of the two approximation schemes is better than the other.
\newline\newline
Figure \ref{fig:DA} shows that both $D_A$ and $z$ individually are not well reproduced by the approximation schemes. A better understanding of the latter can be obtained by recalling equation \eqref{eq:z_exact}: $1+z = e^{\int_{t(\lambda)}^{t_0}dt\left(\frac{1}{3}\Theta + \sigma_{\alpha}^{\beta}e^{\alpha}e_{\beta} \right)}$. The redshift of the covariant scheme is based on the assertion that contributions to the integral from $\sigma_{\alpha}^{\beta}e^{\alpha}e_{\beta}$ and $\Delta\Theta:=\Theta - \left\langle \Theta \right\rangle $ cancel individually in a statistically homogeneous and isotropic universe. In \cite{Tardis,scSZ5} this was found not to be true. Instead, the two terms canceled with each other along the studied light rays. It is clear from figure \ref{fig:DA} that neither type of cancellation happens here. This is shown in detail in figure \ref{fig:comp}. Although the shear and expansion fluctuations tend to contribute to the integrals with different signs, the former contributes much more significantly than the latter.
\newline\indent
Again, the result does not seem to depend on the light rays being propagated radially through the structures as similar plots are obtained for light rays propagated with random impact parameters at turn-around points. The disagreement between the exact and average/approximate redshift-distance relations can therefore not be mitigated (much) by averaging over many light rays with random impact parameters corresponding to light rays propagating in a statistically homogeneous and isotropic universe.
\newline\newline
\begin{figure}
\centering
\subfigure[]{
\includegraphics[scale = 0.48]{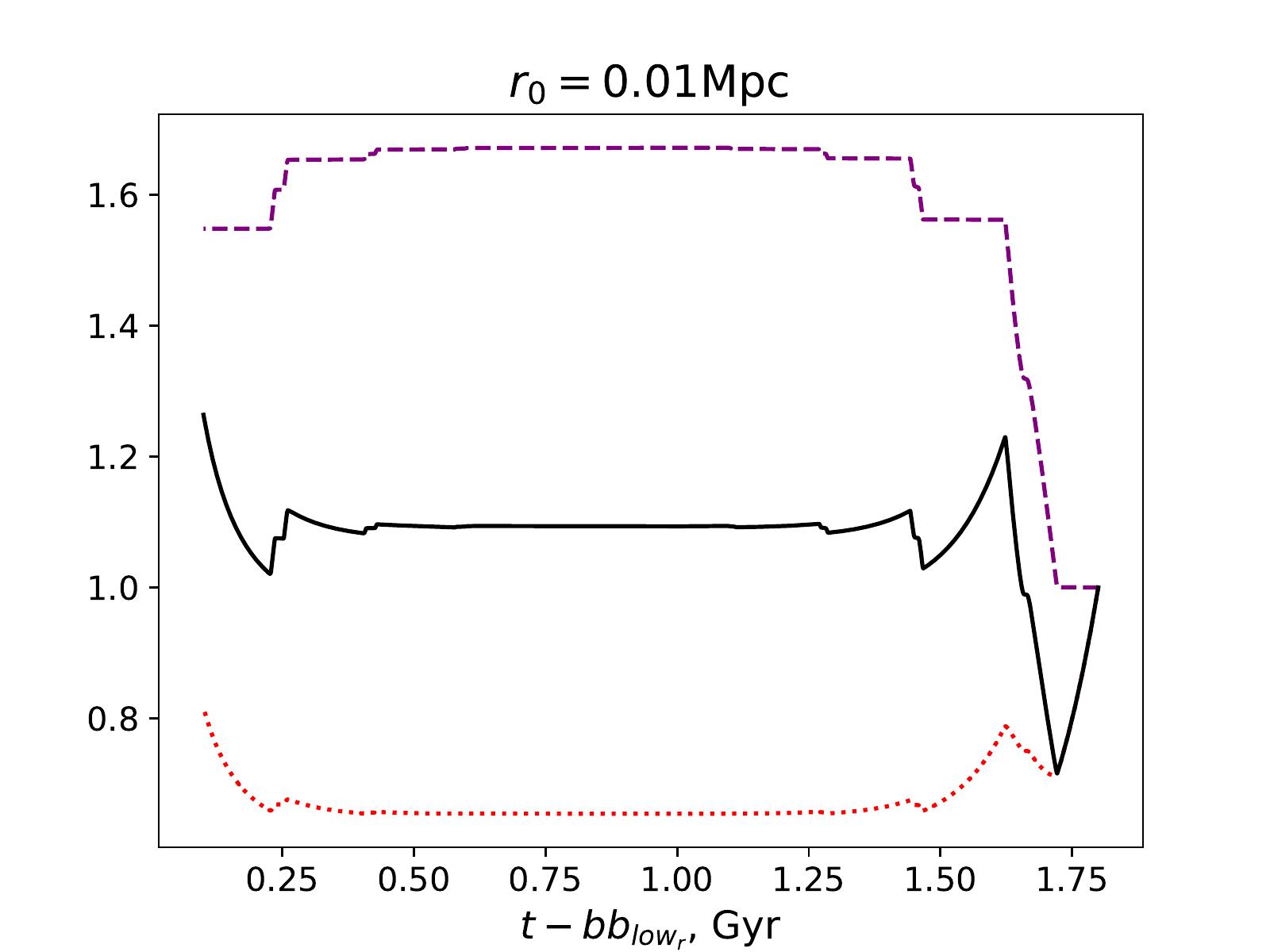}
}
\subfigure[]{
\includegraphics[scale = 0.48]{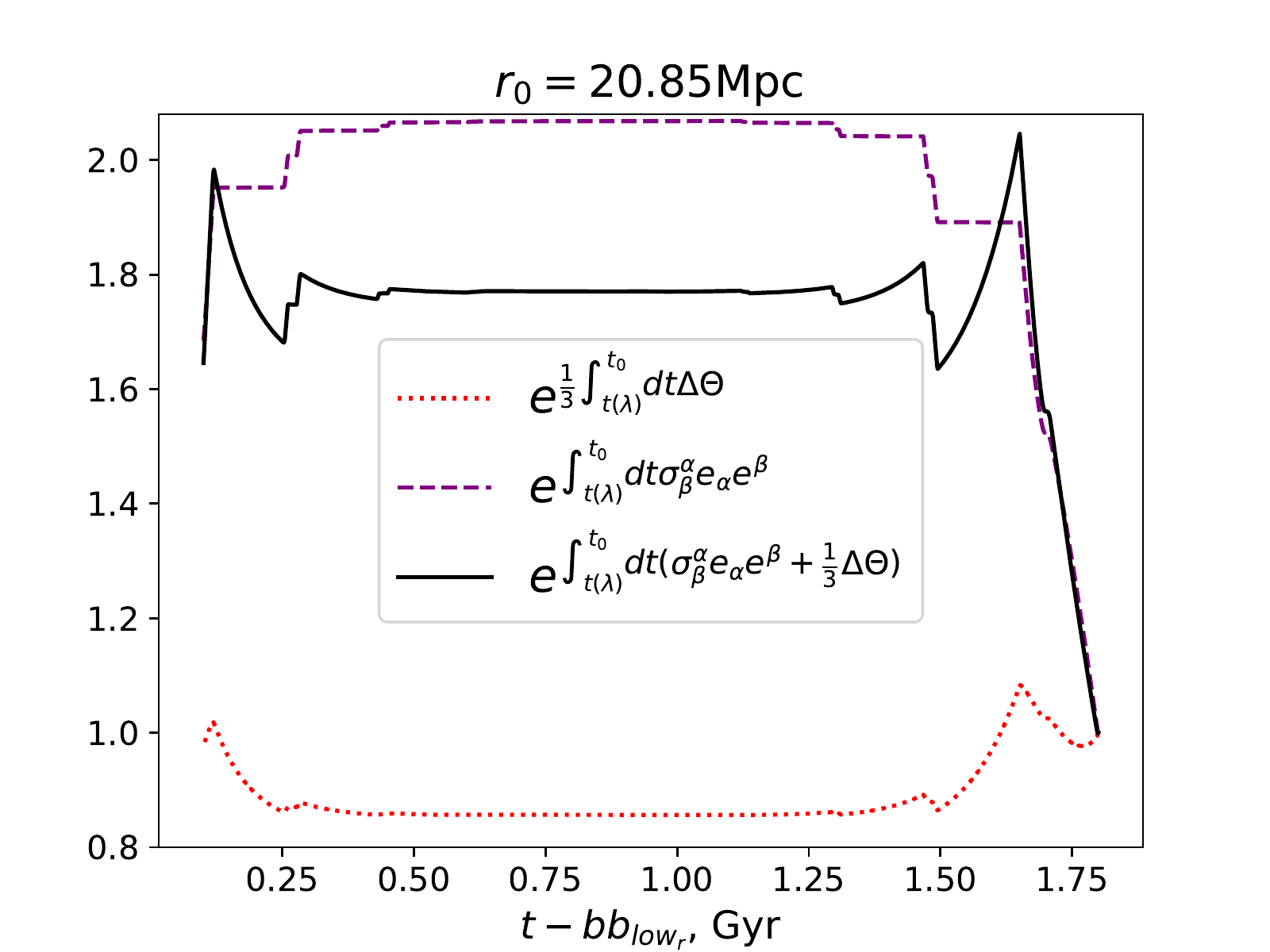}
}
\subfigure[]{
\includegraphics[scale = 0.48]{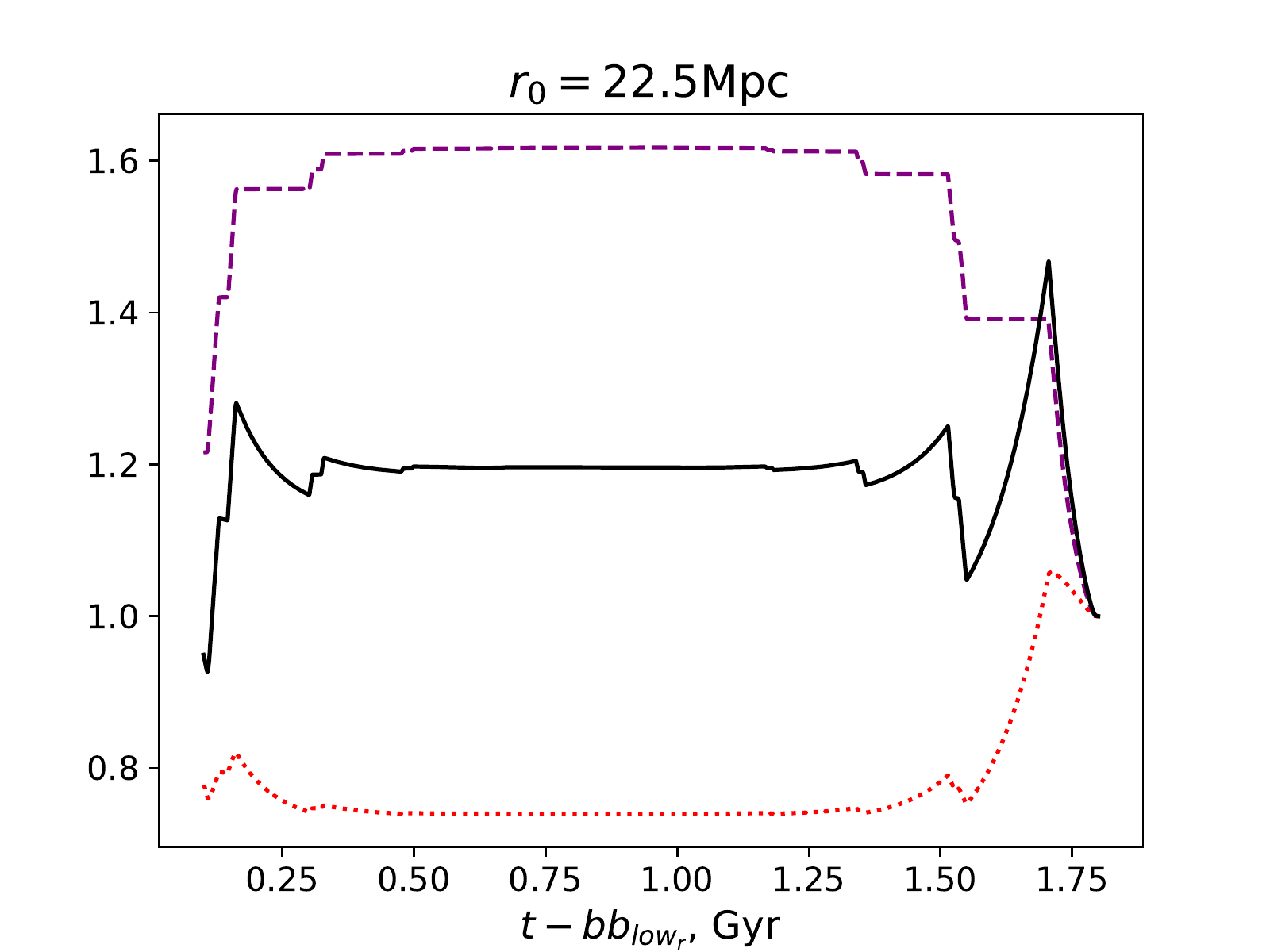}
}
\caption{Components of the redshift along radial light rays with the observer placed at different radial coordinates (indicated at the top of each figure by the value of $r_0$). Labels given in the upper right plot are valid for all three plots. The redshift components are plotted against the time coordinate along the light rays since this is monotonic while the redshift itself is highly non-monotonic.}
\label{fig:comp}
\end{figure}
Figure \ref{fig:comp} shows that the average expansion rate along individual light rays is not approximated well by the volume-averaged expansion rate. This was also found to be the case in \cite{Tardis} but it was there not clear to what extent the result was an artifact of surface layers. In \cite{Tardis} is was also found that the average density along light rays was not well approximated by the volume-averaged density. Figure \ref{fig:rho} shows the density along the three presented light rays divided by the volume-averaged density. As seen, the density distributions along the light rays look somewhat peculiar compared to what one would expect to see in the real universe, emphasizing that the model should not be considered as meant to model the real universe.
\newline\newline
\begin{figure}
\centering
\subfigure[]{
\includegraphics[scale = 0.48]{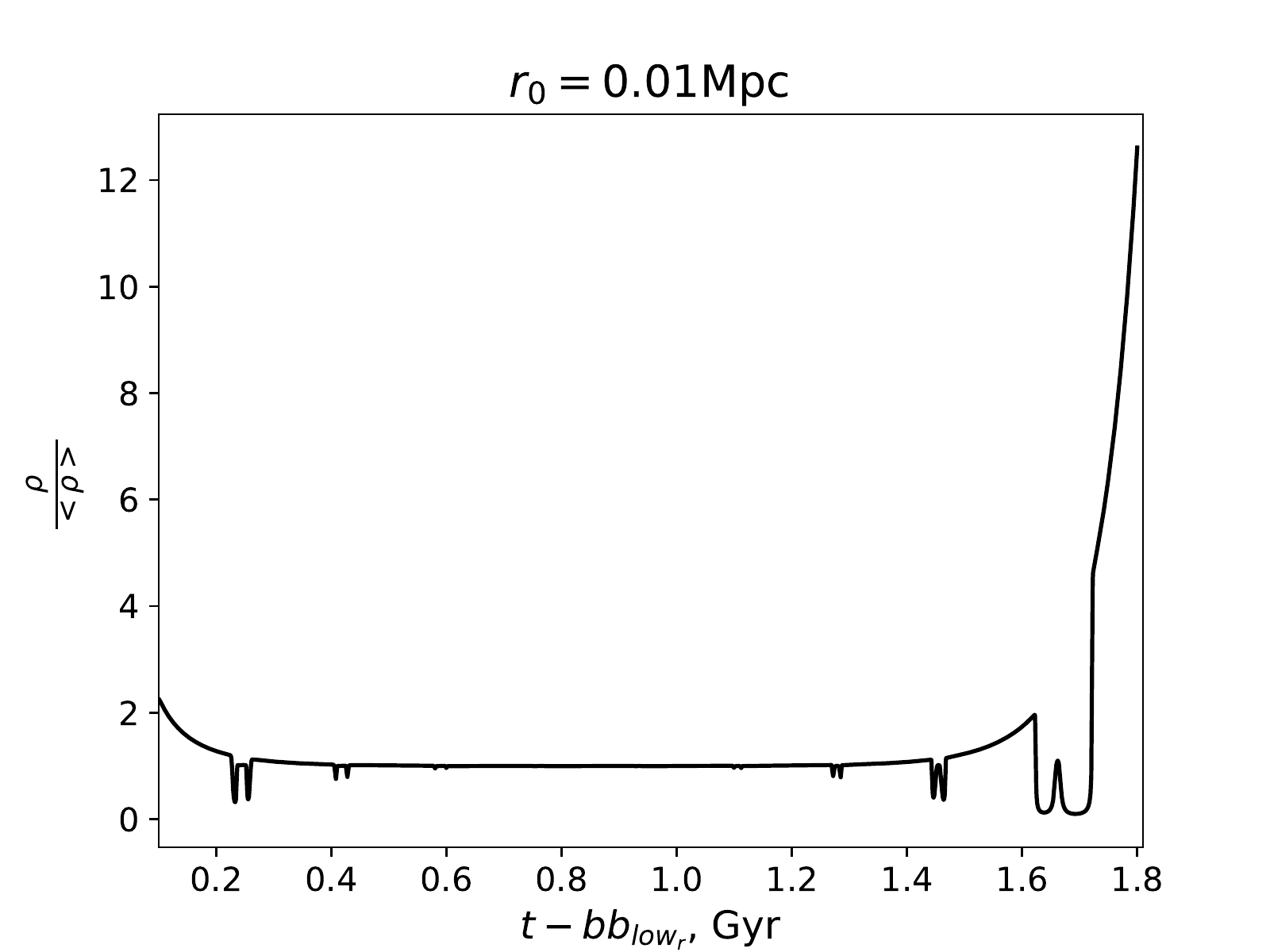}
}
\subfigure[]{
\includegraphics[scale = 0.48]{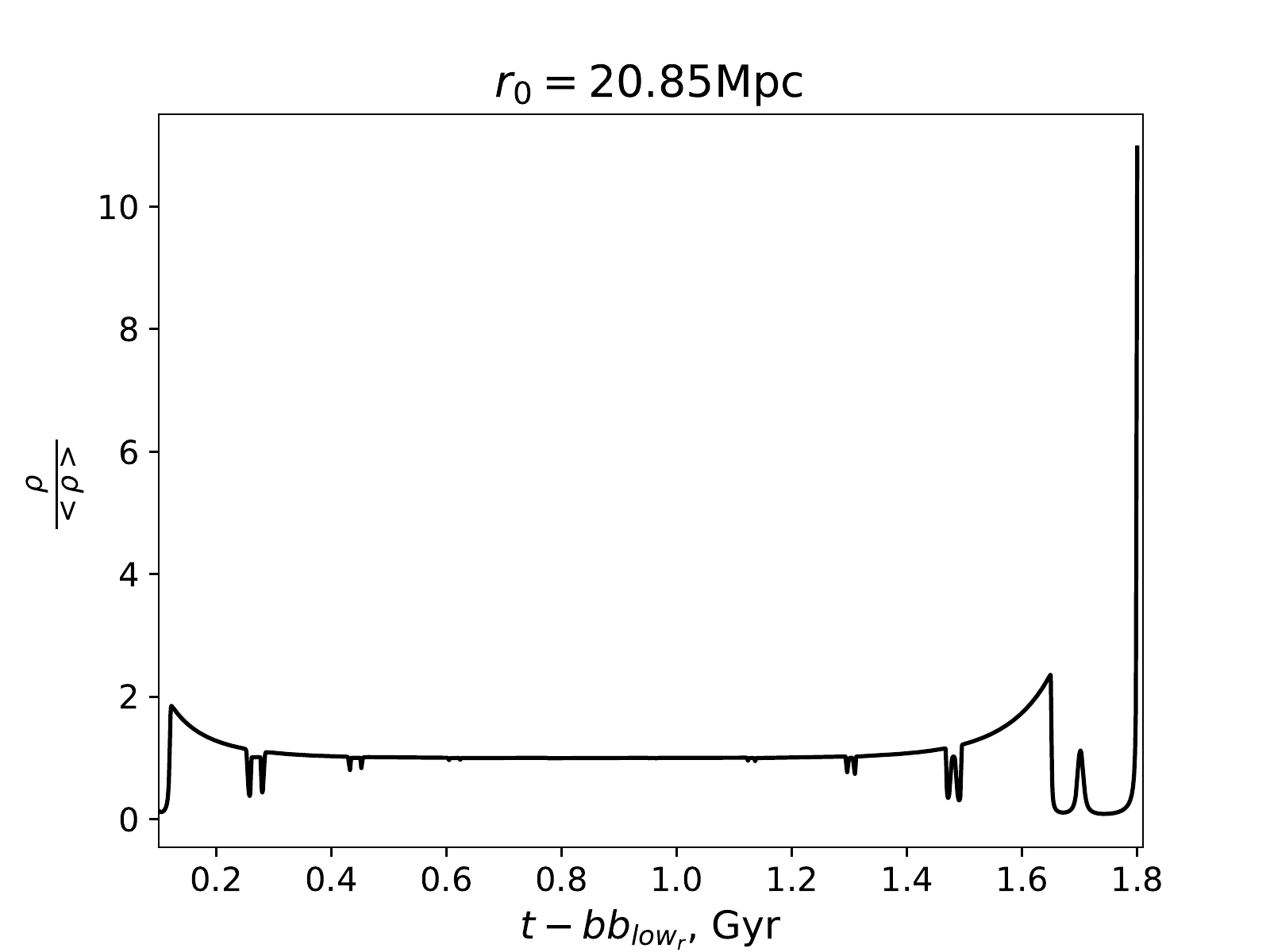}
}
\subfigure[]{
\includegraphics[scale = 0.48]{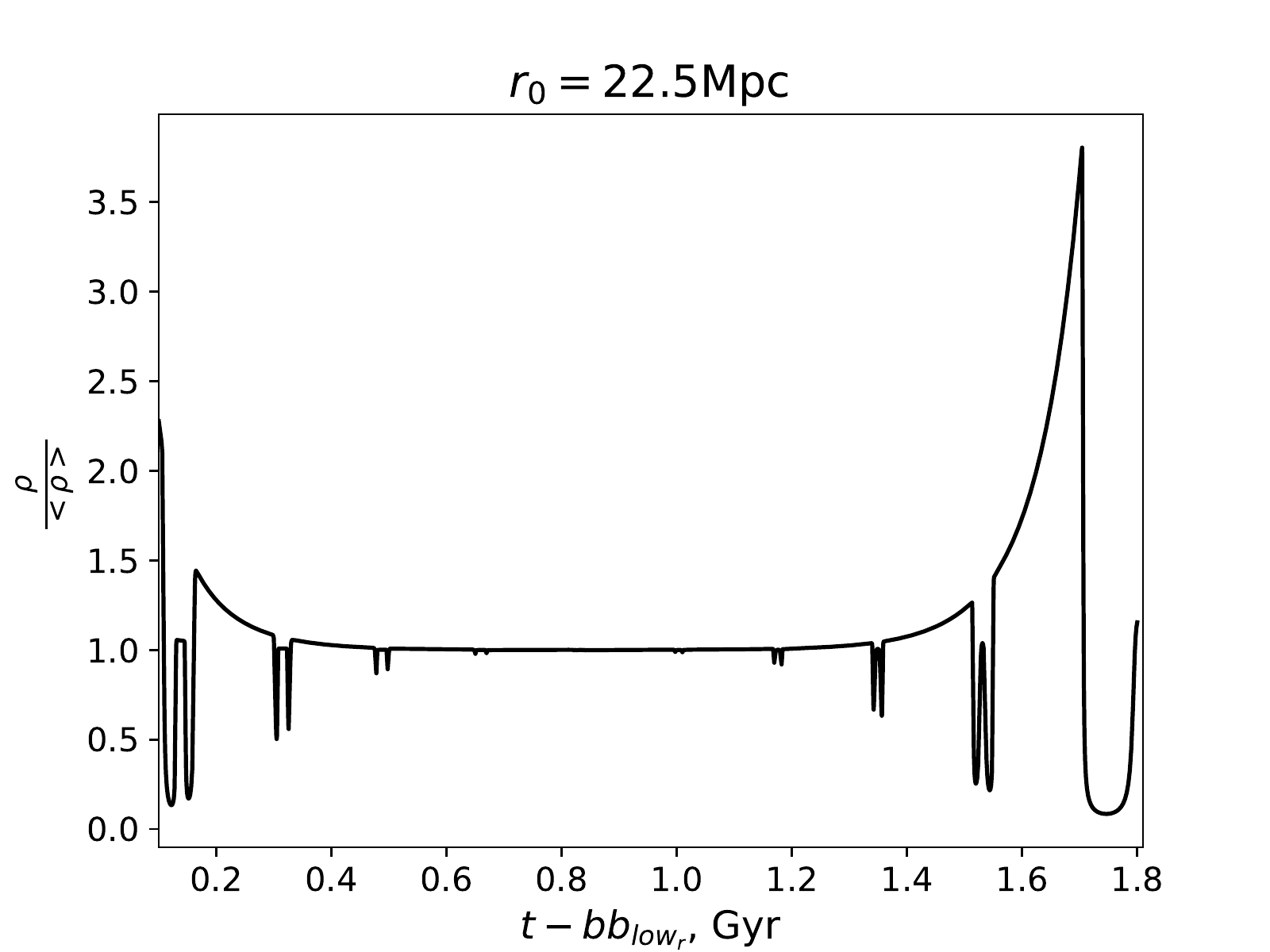}
}
\caption{Density along radial light rays with the observer placed at different radial coordinates (indicated at the top of each figure by the value of $r_0$). The density is plotted against the time coordinate along the light rays since this is monotonic while the redshift itself is highly non-monotonic.}
\label{fig:rho}
\end{figure}

The results presented in this section overall indicate that using the two considered approximation schemes based on volume-averaged quantities does not in general yield good on-average light propagation descriptions. However, the disagreement between the exact results and results based on volume-averaged quantities are presumably exaggerated by large local effects in the expansion rate discussed at the beginning of the section. Specifically, although it was attempted to place the observer at a time where the expansion rate was not too large, the results are at least to some degree a result of this attempt being unsuccessful.

\section{Summary}\label{sec:summary}
A method for constructing perfect fluid Swiss-cheese models with non-negligible backreaction was proposed and illustrated with a simple example model. The particular example model exhibits significant backreaction during some time periods and was used to study the validity of schemes for relating volume averages to observations that have been proposed in the literature. A very poor agreement was obtained between the exact light propagation results and the redshift-distance relations predicted by the schemes based on volume averages. While this may in principle be a genuine effect, the particular example model suffers from several features that are not expected to hold in the real universe. For instance, there is a significant effect from the local position of the observer and local expansion rates are in general quite extreme, either being almost vanishing or very large. The main point with the example model is to illustrate the proposed method: To alter loitering FLRW models with large curvature so that they become inhomogeneous and can be used to fill out the ``holes'' in Swiss-cheese models. Future studies aim at using the method with more complex solutions to the Einstein equation, permitting more realistic constructions where the large curvature is constrained to the inhomogeneous parts of the Swiss-cheese models and where the cheese is free of exotic components or extreme parameter values.
\newline\indent
Although the example model studied here cannot be considered realistic, one particular result regarding light propagation in the model is of general interest and therefore worth highlighting: The fluctuations of the expansion rate and the projected shear do not cancel each other along the light rays. This is interesting as such cancellations were found to occur both in \cite{Tardis} and \cite{scSZ5} which were also based on LTB Swiss-cheese models.

\section{Acknowledgments}
The author thanks Syksy Rasanen for conversations on the topic.
\newline\newline
The author is supported by the Independent Research Fund Denmark under grant number 7027-00019B.

\appendix

\section{Scalar LTB quantities and their spatial averages}\label{app:averages}
With the notation given according to the line element $ds^{2} = -dt^2 +\frac{A^2_{,r}(t,r)}{1-k(r)}dr^2 +A(t,r)^2d\Omega^2$, the expansion rate and spatial curvature of the LTB metric are given by
\begin{equation}
\Theta = \frac{A_{,tr}}{A_r} + 2\frac{A_{,t}}{A}
\end{equation}
\begin{equation}
^{(3)}R = \frac{k}{A^2}+\frac{k_{,r}}{A_{,r}A}
\end{equation}
The volume averages of these quantities and of the matter density as well as $Q$ in a volume $r<r_D$ can be written as
\begin{equation}
V\left\langle \Theta\right\rangle  = 4\pi \int_{0}^{r_D}\frac{A^2dr}{\sqrt{1-k}}\left(A_{,tr} + 2\frac{A_{,t}A_{,r}}{A} \right) 
\end{equation}
\begin{equation}
V\left\langle^{(3)}R \right\rangle = 8\pi\int_{0}^{r_D}\frac{dr}{\sqrt{1-k}}\left(A_{,r}k + ak_{,r} \right) 
\end{equation}
\begin{equation}
V\left( Q +\frac{2}{3}\left\langle \Theta\right\rangle ^2\right) = 8\pi \int_{0}^{r_D}\frac{dr}{\sqrt{1-k}}\left(A_{,t}^2A_{,r} + 2A_{,t}A_{,tr}A \right) 
\end{equation}
\begin{equation}
V\left\langle \rho\right\rangle  = \frac{1}{G_N}\int_{0}^{r_D}\frac{M_{,r}dr}{\sqrt{1-k}}.
\end{equation}
Notice that the left hand sides of the above averages are multiplied by the proper volume $V$ of the averaging domain.

\section{Riemann components for LTB metric}\label{app:light}
With the LTB metric written according to the line element $ds^2 = -dt^2 + Rdr^2+Fd\theta^2 + Pd\theta^2$, the non-vanishing Riemann components up to symmetries are:
\begin{equation}
\mathcal{R}_{rtrt} = -\frac{ 2RR_{,tt}-R_{,t}^2 }{4R}
\end{equation}
\begin{equation}
\mathcal{R}_{\theta t\theta t} = -\frac{2FF_{,tt}-F_{,t}^2}{4F}
\end{equation}
\begin{equation}
\mathcal{R}_{t\phi t\phi } = -\frac{2PP_{,tt}-P_{,t}^2}{4P}
\end{equation}
\begin{equation}
\mathcal{R}_{r\theta r\theta} = \frac{FRF_{,t}R_{,t}+c^2FF_{,r}R_{,r}+Rc^2\left(F_{,r}^2-2FF_{,rr} \right) }{4c^2FR}
\end{equation}
\begin{equation}
\mathcal{R}_{r \phi r \phi} = \frac{PRP_{,t}R_{,t}+c^2PP_{,r}R_{,r} + c^2R\left(P_{,r}^2-2PP_{,rr} \right) }{4c^2PR}
\end{equation}
\begin{equation}
\mathcal{R}_{\theta\phi\theta\phi} = -\frac{c^2PF_{,r}P_{,r}+R\left( 2c^2PP_{,\theta\theta} - c^2P_{,\theta}^2 - PF_{,t}P_{,t} \right) }{4c^2PR}
\end{equation}
$\mathbf{F}$ thus becomes:
\begin{equation}
\begin{split}
-2F = \mathcal{R}_{trtr}\left[ \left(\epsilon^r \right) ^2\left( k^t\right) ^2 + \left(\epsilon^t \right) ^2\left( k^r\right) ^2 - 2\epsilon^t\epsilon^r k^t k^r \right] +\\
\mathcal{R}_{\theta\phi\theta\phi}\left[ (\epsilon^\phi)^2(k^{\theta})^2 + (\epsilon^{\theta})^2(k^{\phi})^2 -2k^{\theta}k^{\phi}\epsilon^{\theta}\epsilon^{\phi}  \right]+\\
\mathcal{R}_{t\theta t \theta}[ \left(\epsilon^t \right) ^2\left\lbrace \left( k^{\theta} \right) ^2 + \sin^2(\theta)(k^{\phi})^2 \right\rbrace  + (k^t)^2\left\lbrace (\epsilon^{\theta})^2 +\sin^2(\theta)(\epsilon^{\phi})^2\right\rbrace -\\
2(\epsilon^t)k^t\left\lbrace \epsilon^{\theta}k^{\theta} + \sin^2(\theta)\epsilon^{\phi}k^{\phi}   \right\rbrace ] + \\
\mathcal{R}_{r\theta r\theta}[ (\epsilon^r)^2\left\lbrace (k^{\theta})^2 + (k^{\phi})^2\sin^2(\theta)\right\rbrace +\\
(k^r)^2\left\lbrace (\epsilon^{\theta})^2 + (\epsilon^{\phi})^2\sin^2(\theta) \right\rbrace -2\epsilon^r k^r \left\lbrace \epsilon^{\theta}k^{\theta} + \sin^2(\theta)\epsilon^{\phi}k^{\phi} \right\rbrace   ]   
\end{split}
\end{equation}

\end{document}